\documentclass[pre,preprint,english,aps,prb,showpacs,a4paper,floatfix]{revtex4}
\usepackage[T1]{fontenc}
\usepackage{color}
\usepackage{amsmath}
\usepackage{amssymb}
\usepackage[latin1]{inputenc}
\usepackage{graphicx}
\usepackage{bm}
\usepackage{epsfig}

\begin{document}

\title{Effect of shape biaxiality on the phase behavior of colloidal liquid-crystal monolayers}

\author{Miguel Gonz\'alez-Pinto}
\email{miguel.gonzalezp@uam.es}
\affiliation{Departamento de F\'{\i}sica Te\'orica de la Materia Condensada, Facultad de Ciencias,
Universidad Aut\'onoma de Madrid, E-28049 Madrid, Spain}

\author{Yuri Mart\'{\i}nez-Rat\'on}
\email{yuri@math.uc3m.es}
\affiliation{Grupo Interdisciplinar de Sistemas Complejos (GISC), Departamento de Matem\'aticas, Escuela Polit\'ecnica Superior,
Universidad Carlos III de Madrid, Avenida de la Universidad 30, E-28911, Legan\'es, Madrid, Spain}

\author{Enrique Velasco}
\email{enrique.velasco@uam.es}
\affiliation{Departamento de F\'{\i}sica Te\'orica de la Materia Condensada, Instituto de Ciencia de Materiales Nicol\'as
Cabrera and Condensed Matter Physics Center (IFIMAC), Facultad de Ciencias, Universidad Aut\'onoma de Madrid, E-28049 Madrid, Spain}

\author{Szabolcs Varga}
\email{vargasz@almos.uni-pannon.hu}
\affiliation{Institute of Physics and Mechatronics, University of Pannonia, PO Box 158, Veszpr\'em, H-8201 Hungary}

\begin{abstract}
We extend our previous work on monolayers of uniaxial particles [J. Chem. Phys. {\bf 140}, 204906 (2014)]
to study the effect of particle biaxiality on the phase behavior of liquid-crystal monolayers. Particles are modelled as 
board-like hard bodies with three different edge lengths $\sigma_1\geq\sigma_2\geq\sigma_3$, and use is made of
the restricted-orientation approximation (Zwanzig model). A density-functional formalism based on the fundamental-measure theory 
is used to calculate phase diagrams for a wide range of values of the largest aspect ratio ($\kappa_1=\sigma_1/\sigma_3\in[1,100]$).
We find that particle biaxiality in general destabilizes the biaxial nematic phase already present in monolayers of uniaxial particles. 
While plate-like particles exhibit strong biaxial ordering, rod-like ones with $\kappa_1>21.34$ exhibit reentrant uniaxial and biaxial phases.
As particle geometry is changed from uniaxial- to increasingly
biaxial-rod-like, the region of biaxiality is reduced, eventually ending in a critical-end point. 
For $\kappa_1>60$, a density gap opens up in which the biaxial nematic phase is stable for any particle biaxiality. 
Regions of the phase diagram where packing-fraction inversion occurs (i.e. packing fraction is a decreasing function of density) 
are found. Our results are compared with the recent experimental studies on nematic phases of magnetic nanorods.    
\end{abstract}

\date{\today}


\maketitle


\section{Introduction}

Biaxial hard-particle systems have received considerable theoretical and experimental attention since their first theoretical 
prediction by Freiser \cite{1}. The characteristic feature of the biaxial phase is that two directions of orientational
ordering occur associated with two molecular symmetry axes \cite{2}.
The importance of studying biaxial nematic 
phases is that they might be used in practical applications, such as fast electro-optical devices \cite{3}.  

The theoretical exploration of stable biaxial nematic order has been based on biaxial hard-body and Gay-Berne-type soft 
potential models, using specific particle shapes such as spheroplatelets \cite{4,5}, biaxial ellipsoids \cite{6,7} and bent-core 
particles \cite{8,9,10,11}. 
The biaxial nematic phase has also been found in binary mixtures of uniaxial plate-like and rod-like particles \cite{12,13,14,15,16,17}. 
However, the experimental realization of this exotic phase has proved to be rather complicated. The first observation
dates back to the study of Yu and Saupe \cite{18}, who observed that a mixture of potassium laurate, 1-decanol, 
and water exhibited a region of biaxial order between two uniaxial phases. Later, biaxial nematic order was observed in low molecular 
weight thermotropic liquid crystals where the constituting particles had biaxial symmetry \cite{19,20}. Regarding the shape 
of the constituting particles, banana-shaped mesogenic molecules are found to form thermotropic biaxial nematic phase \cite{21,22}, 
while board-shaped colloidal particles have been used successfully in the stabilization of lyotropic biaxial nematic phases 
\cite{23,24,25}. 

The recent experimental observation of biaxial nematic order in suspensions of board-like goethite nanorods \cite{23,24,25}
has prompted several 
theoretical studies in order to determine the global phase behavior of hard board-shaped particles \cite{5,26} and 
also to identify those processes 
which promote the formation of the biaxial nematic phase \cite{27,28}. Interestingly, an increasing polydispersity in shape 
and size favors the biaxial nematic phase over other ordered phases \cite{27}. In addition to this, binary mixtures consisting
of board-shaped particles with added polymers can stabilize biaxial order very efficiently \cite{28}. Even the biaxiality 
of the nematic phase can be tuned by applying an external magnetic field \cite{29}. 

By inserting goethite nanorods into a soft lamellar matrix of non-ionic surfactant, it is also possible to examine the effect of dimensional 
reduction on the stability of mesophases \cite{30,31,32}. The confined nanorods between the bilayers of a lamellar phase have been
shown to undergo a first-order in-plane (two-dimensional) isotropic-nematic phase transition, where the isotropic and 
nematic phases correspond to planar and biaxial nematic phases, respectively. In the light of increasing amount of knowledge about 
the ordering properties of board-shaped goethite nanorods in confined geometries, it is worth studying the phase behaviour of hard 
board-shaped particles in quasi-two-dimensions using theoretical methods, and this is the motivation of our work. 

In the present study we use density-functional theory in the fundamental-measure version to examine the orientational and 
positional ordering properties of confined hard-board colloidal particles with discrete orientations.
The confinement is such that the centers of the board particles are 
always on a flat surface. We mainly focus on the effect of shape biaxiality on the stability of the biaxial nematic phase, but we 
also determine the stability regions of other mesophases such as the uniaxial nematic and positionally-ordered smectic, columnar and 
solid phases using bifurcation analysis. An important result is that an increasing biaxiality does not promote the formation 
of biaxial nematic phases due to the free-volume maximizing effect of the packing entropy.

The paper is organized as follows. The particle model and expressions for the relevant
order parameters measuring biaxial ordering are presented in Sec. \ref{II}.
Sec. \ref{III} presents the results, which include the evolution of the phase diagrams with particle 
biaxiality and the density dependence of 
the order parameters for different particle shapes. Some conclusions are drawn in Sec. \ref{IV}. 
Details on the density-functional theory and bifurcation analysis are presented in the Appendices.

\begin{figure}
\includegraphics[width=11cm]{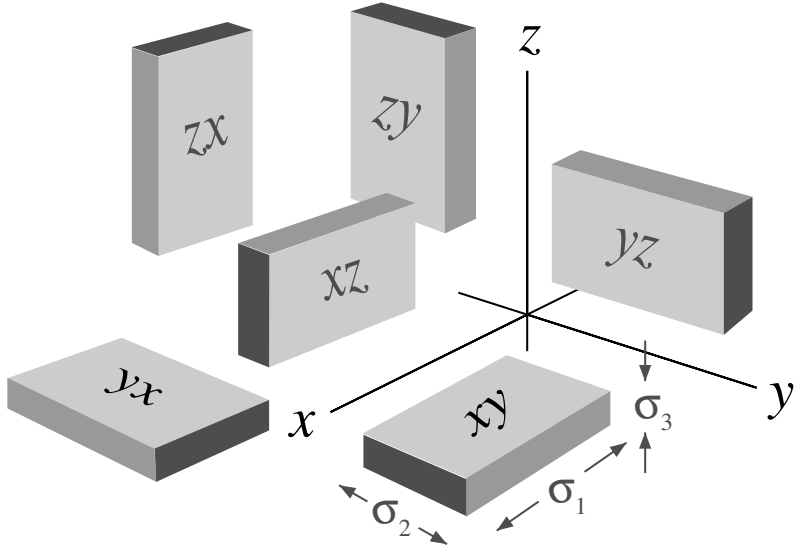}\\
\caption{Six Zwanzig species $\mu\nu$ (with $\mu,\nu=x,y,z$ and $\mu\neq \nu$) of hard board-like particles 
of dimensions $\sigma_1\geq\sigma_2\geq\sigma_3$.}
\label{sketch0}
\end{figure}

\section{Model and Theory}
\label{II}

\begin{figure}
\includegraphics[width=10cm]{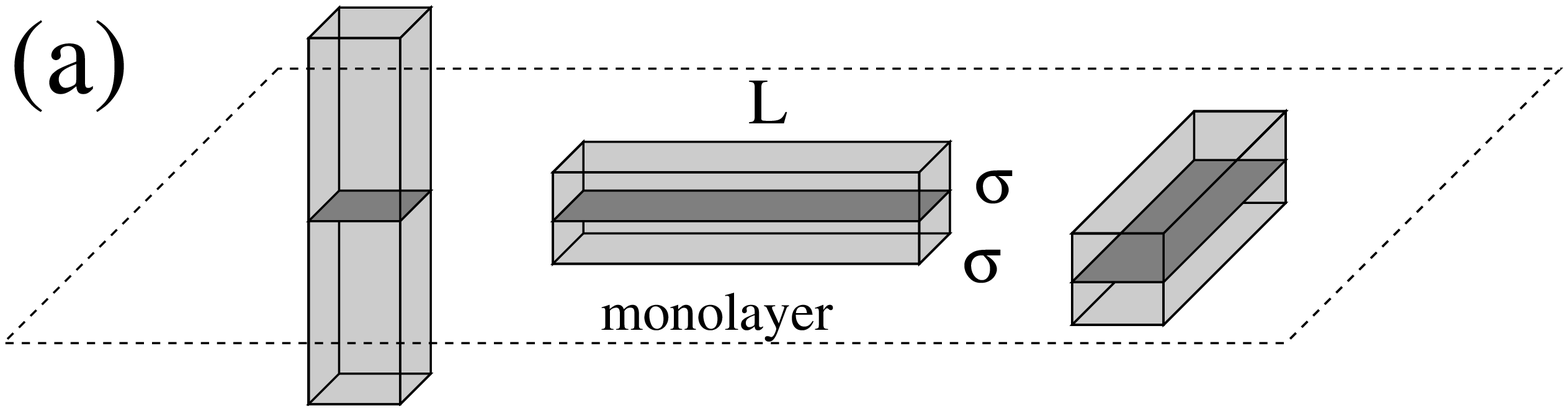}\\
\includegraphics[width=10cm]{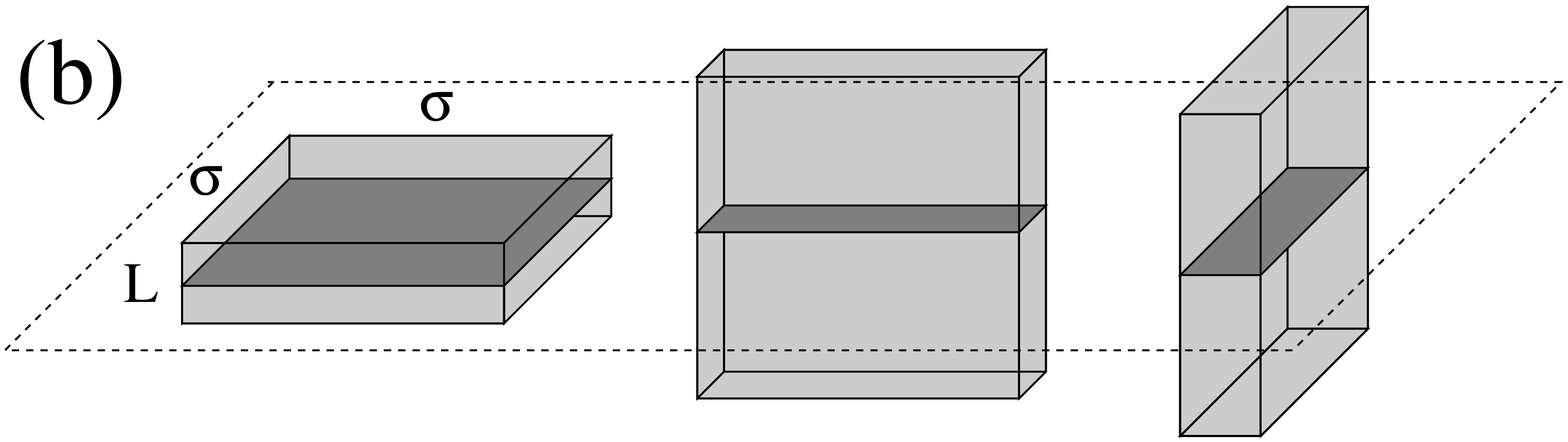}
\caption{Projection of orientation-restricted uniaxial hard boards on the monolayer. 
(a) Uniaxial prolate particles ($\sigma_1=L$ and $\sigma_2=\sigma_3=\sigma$). (b) Uniaxial oblate 
particles ($\sigma_1=\sigma_2=\sigma$, $\sigma_3=L$). The projected areas are conveniently shaded. }
\label{sketch}
\end{figure}

Colloidal particles are modelled as biaxial hard boards with edge-lengths
$\sigma_1\geq\sigma_2\geq\sigma_3$ and centres of mass located on a flat surface perpendicular to the $z$ axis.
Particles are allowed to rotate (within the restricted-orientation approximation) in the full 3D solid angle, 
but constrained to move on a plane. By restricting the possible orientations to be the three Cartesian axes, and
considering the symmetries of the particles, six possible orientations, depicted in Fig. \ref{sketch0}, are 
possible. 
The system can then be mapped onto a six-component mixture, with species labelled by $\mu\nu$ (with $\mu,\nu=x,y,z$ 
and $\mu\neq\nu$), where the indexes refer to the orientation of the longest and intermediate particle lengths, 
respectively. The density of `species' $\mu\nu$ is written as $\rho_{\mu\nu}=\rho \gamma_{\mu\nu}$, with $\rho$ 
the 2D total density. $\{\gamma_{\mu\nu}\}$ is a set of molar fractions that fulfills the constraint 
$\displaystyle\sum_{\mu,\nu}\gamma_{\mu\nu}=1$. The particular cases of prolate ($\sigma_1=L$ and $\sigma_2=\sigma_3=\sigma$) 
and oblate ($\sigma_1=\sigma_2=\sigma$ and $\sigma_3=L$) particles are sketched in Figs. \ref{sketch}(a) 
and (b), respectively.

To characterise particle shape, two aspect ratios are defined, $\kappa_1=\sigma_1/\sigma_3$ and $\kappa_2=\sigma_2/\sigma_3$,
which fulfill the inequalities $1\leq \kappa_2\leq \kappa_1$. Further, the degree of particle biaxiality will be
characterised by the parameter
\begin{eqnarray}
\theta\equiv (\kappa_1-1)^{-1}\left(\frac{\kappa_1}{\kappa_2}-\kappa_2\right).
\end{eqnarray}
For fixed $\kappa_1$, the $\theta$ parameter varies from $-1$ (when $\kappa_2=\kappa_1$, corresponding to 
uniaxial plate-like geometry) to $\theta=1$ (when $\kappa_2=1$, pertaining to uniaxial rod-like geometry).  
The value $\theta=0$ corresponds to perfect biaxiality, i.e. $\kappa_2=\sqrt{\kappa_1}$; 
when $\kappa_2\gtrless\sqrt{\kappa_1}$ particles are considered to be oblate or prolate, respectively. 

The statistical mechanics of the monolayer is dealt with using a version of density-functional theory.
This version is based on the fundamental-measure theory for hard cubes \cite{Cuesta}. The resulting free-energy functional 
is expressed as a function of the set of molar fractions $\{\gamma_{\mu\nu}\}$, and the equilibrium state of the monolayer is 
obtained by minimising the free energy with respect to this set with the constraint 
$\displaystyle\sum_{\mu,\nu}\gamma_{\mu\nu}=1$ and for fixed scaled density $\rho^*=\rho\sigma_3^2$, where $\rho=N/A$ is the
two-dimensional density ($N$ is number of particles and $A$ the area of monolayer).
The chemical potentials $\mu_{\tau\nu}$ of all species and the lateral pressure $p$ of the monolayer can then be
calculated, and phase equilibria can be obtained. Details on the density functional used are given in Appendix
\ref{apen1}. The stability analysis and bifurcation theory derived to obtain the biaxial nematic
spinodal and the nematic stability against non-uniform fluctuations can be found in Appendices \ref{apen2}
and \ref{apen3}.

A useful measure of the ordering properties of the equilibrium phases are the order parameters, which help identify
the two possible nematic phases in our system: the uniaxial nematic phase, N$_{\rm u}$, and the biaxial nematic phase,
N$_{\rm b}$. In the case of biaxial particles two order parameter tensors can be defined, 
\begin{eqnarray}
\hat{Q}_{\alpha\beta}=\frac{1}{2}\left(3\langle u_{\alpha}u_{\beta}\rangle -\delta_{\alpha\beta}\right),\hspace{0.6cm}
\hat{B}_{\alpha\beta}=\frac{1}{2}\left(\langle n_{\alpha}n_{\beta}\rangle-\langle m_{\alpha}m_{\beta}\rangle\right),
\end{eqnarray}
where $u_{\alpha}$, 
$n_{\alpha}$ and $m_{\alpha}$ are the $\alpha$-components of the unit vectors ${\bm u}$, ${\bm n}$ and ${\bm m}$ 
along the longest, intermediate and smallest particle lengths. Averages are taken over the orientational distribution 
function, given by the set $\{\gamma_{\mu\nu}\}$. For our restricted-orientation approximation, it can easily be 
shown that the tensors are diagonal:
\begin{eqnarray}
\hat{Q}=
\begin{pmatrix}
\displaystyle{-\frac{Q-\Delta_Q}{2}} & 0 & 0\\
0 & \displaystyle{-\frac{Q+\Delta_Q}{2}} & 0\\
0 & 0 & Q
\end{pmatrix},
\quad \hat{B}=
\begin{pmatrix}
\displaystyle{-\frac{B-\Delta_B}{2}} & 0 & 0\\
0 & \displaystyle{-\frac{B+\Delta_B}{2}} & 0\\
0 & 0 & B
\end{pmatrix}
\end{eqnarray}
where $Q$ and $B$ are uniaxial nematic order parameters,
\begin{eqnarray}
Q\equiv Q_{zz}=\frac{1}{2}\left(3\sum_{\nu\neq z} \gamma_{z\nu}-1\right), \label{la_Q}\quad
B\equiv B_{zz}=\frac{1}{2}\left(\sum_{\mu\neq z}\gamma_{\mu z}-\sum_{\mu,\nu\neq z}\gamma_{\mu\nu}\right),
\label{la_S}
\end{eqnarray}
with $Q$ the usual uniaxial order parameter (note that $B\neq 0$ for both N$_{\rm u}$ and N$_{\rm b}$ phases), 
while    
\begin{eqnarray}
&&\Delta_Q\equiv Q_{xx}-Q_{yy}=\frac{3}{2}\left(\sum_{\nu\neq x}\gamma_{x\nu}-\sum_{\nu\neq y}\gamma_{y\nu}\right),
\label{la_delta}\\
&&\Delta_B\equiv B_{xx}-B_{yy}=\frac{1}{2}\left[\sum_{\mu\neq x}\gamma_{\mu x}-\sum_{\mu\neq y}\gamma_{\mu y}
+\sum_{\mu,\nu\neq y}\gamma_{\mu \nu}-\sum_{\mu,\nu\neq x}\gamma_{\mu \nu}\right],\label{la_Delta}
\end{eqnarray}
are biaxial nematic order parameters, both different from zero only for the N$_{\rm b}$ phase. 

A comment on the definition of the above order parameters in relation with the particle geometry is in order.
For uniaxial rods ($\theta=1$), the vector ${\bm u}$ points along the main symmetry axis (longest particle length), 
the other two being equivalent. Thus, the above definitions for $\{Q,B,\Delta_Q,\Delta_B\}$ are correct in the 
limit $\theta\to 1$, and they will be used for any $\theta>0$. However, for uniaxial oblate particles ($\theta=-1$), 
the main particle axis should be taken to lie along the shortest particle length ${\bm m}$, the other two being 
equivalent: ${\bm u}$ and ${\bm m}$ should be interchanged for $\theta<0$, and all four order parameter
can be obtained from the same formulas as before but replacing $\gamma_{\mu\nu}$ by $\gamma_{\tau\nu}$ 
(with $\tau\neq\mu$).  
In this way we obtain, for example, that $\displaystyle{Q\to-\frac{1}{2}}$ for perfect 
planar nematic ordering, as it should be. 

In the following, the parameter $\Delta_Q^*\equiv 2\Delta_Q/3$ will be used to measure 
the degree of biaxiality for uniaxial plate-like and rod-like particles ($\theta=\pm 1$), while in the case of  
biaxial particles $(-1<\theta<1)$ the parameter $\Delta_{B}$ will be used. 
It can be shown that, for perfect biaxial order, $|\Delta_B|\to 1$ and $0.5$ for rods and plates, respectively, 
while $|\Delta^*_Q|\to 1$ for both particles. In any case, we always plot absolute values of biaxial order parameters
in the figures.

Finally, a useful measure of packing in the monolayer is $\eta$, the area fraction covered by particles on the monolayer 
(packing fraction). It can be shown that $\eta$ is related to $\rho^*$, $Q$ and $B$ by 
\begin{eqnarray}
\eta=\frac{\rho^*}{3}\left\{\kappa_1(\kappa_2+1)+\kappa_2-\left[\kappa_1(\kappa_2+1)-2\kappa_2\right]Q
-3\kappa_1(\kappa_2-1)B\right\}.
\label{la_packing}
\end{eqnarray}
This equation is used later to explain packing-fraction inversion effects that take place for some particle symmetries.

\section{Results}
\label{III}

\begin{figure}
\includegraphics[width=8.15cm]{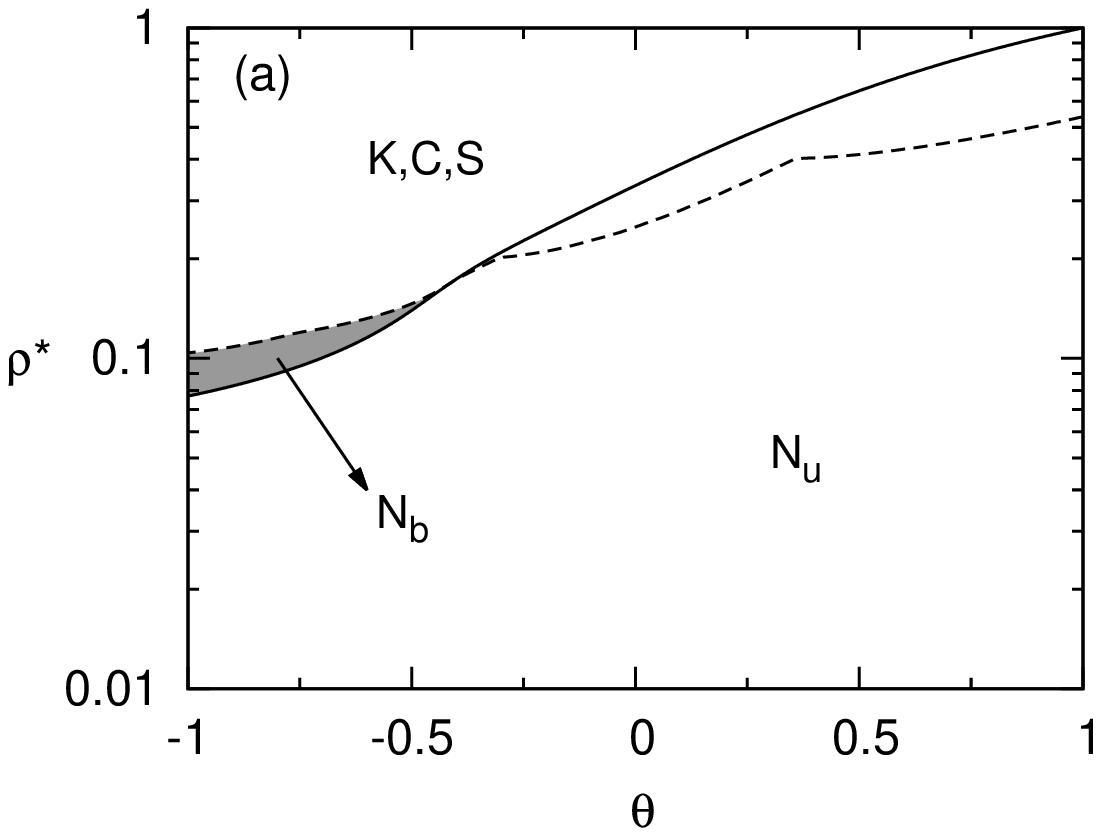}
\includegraphics[width=8.15cm]{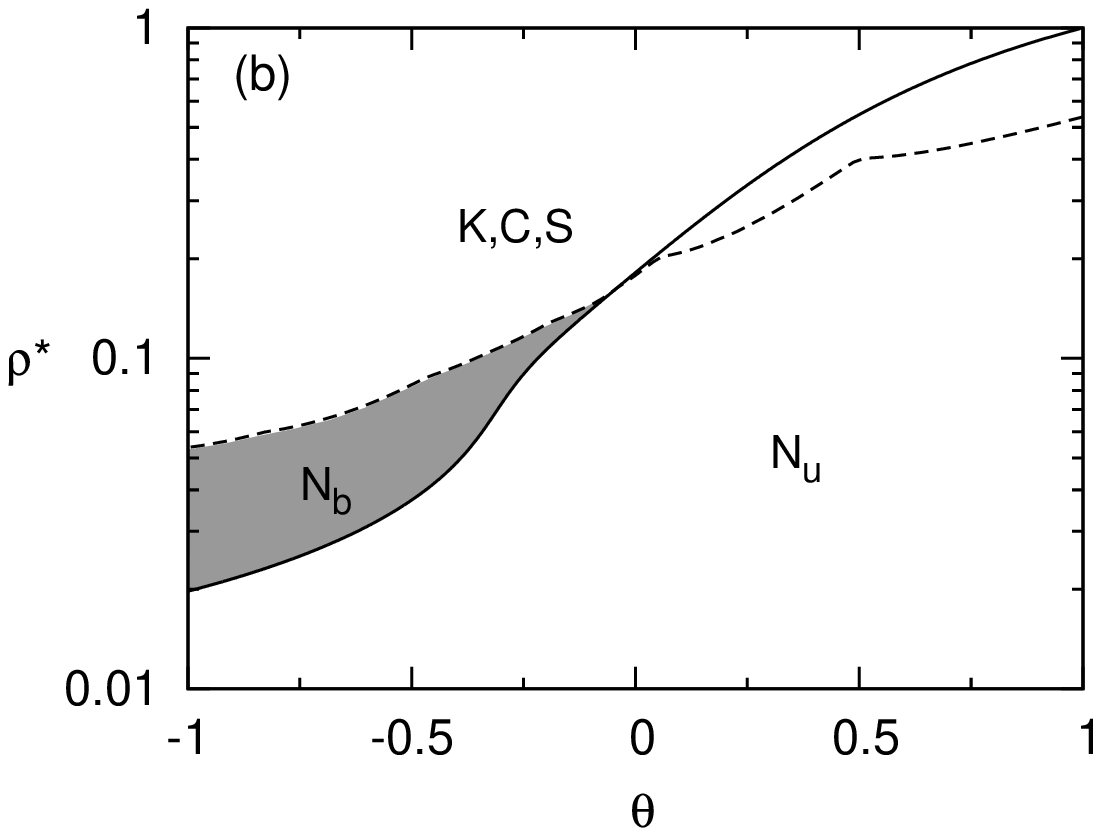}
\includegraphics[width=8cm]{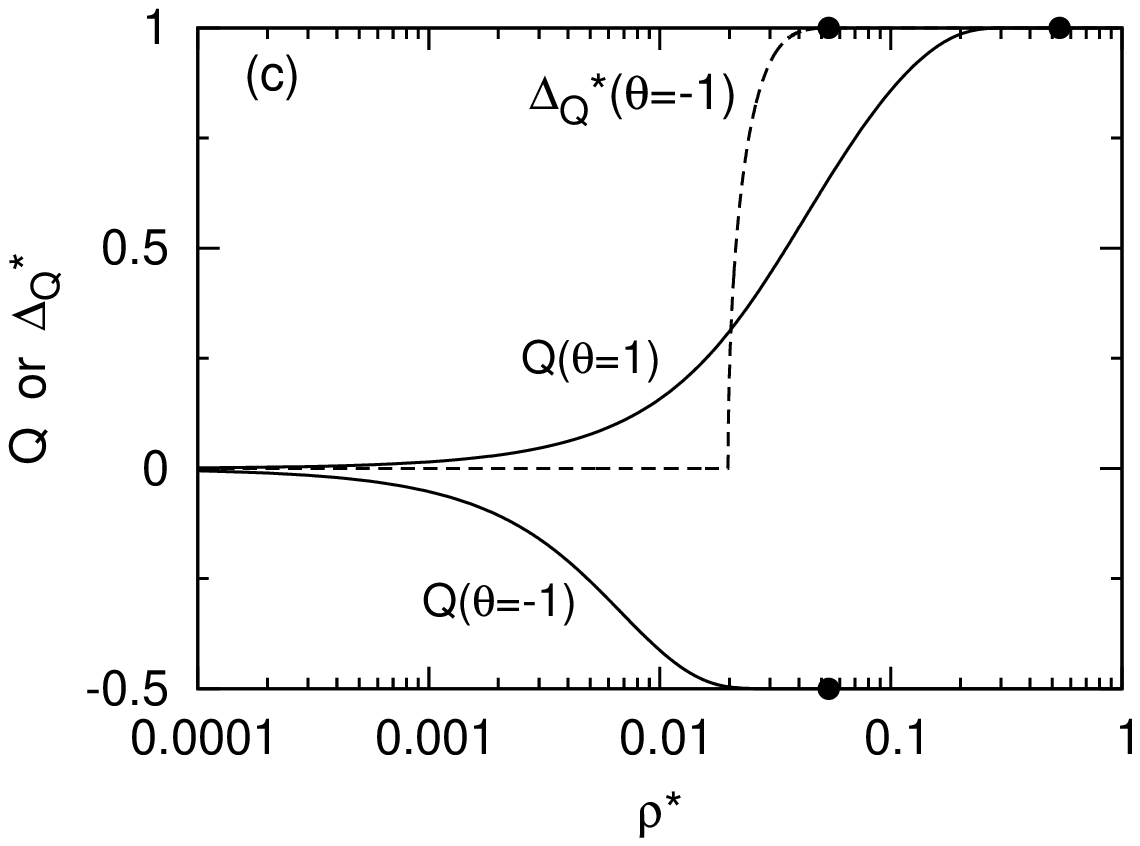}
\includegraphics[width=8cm]{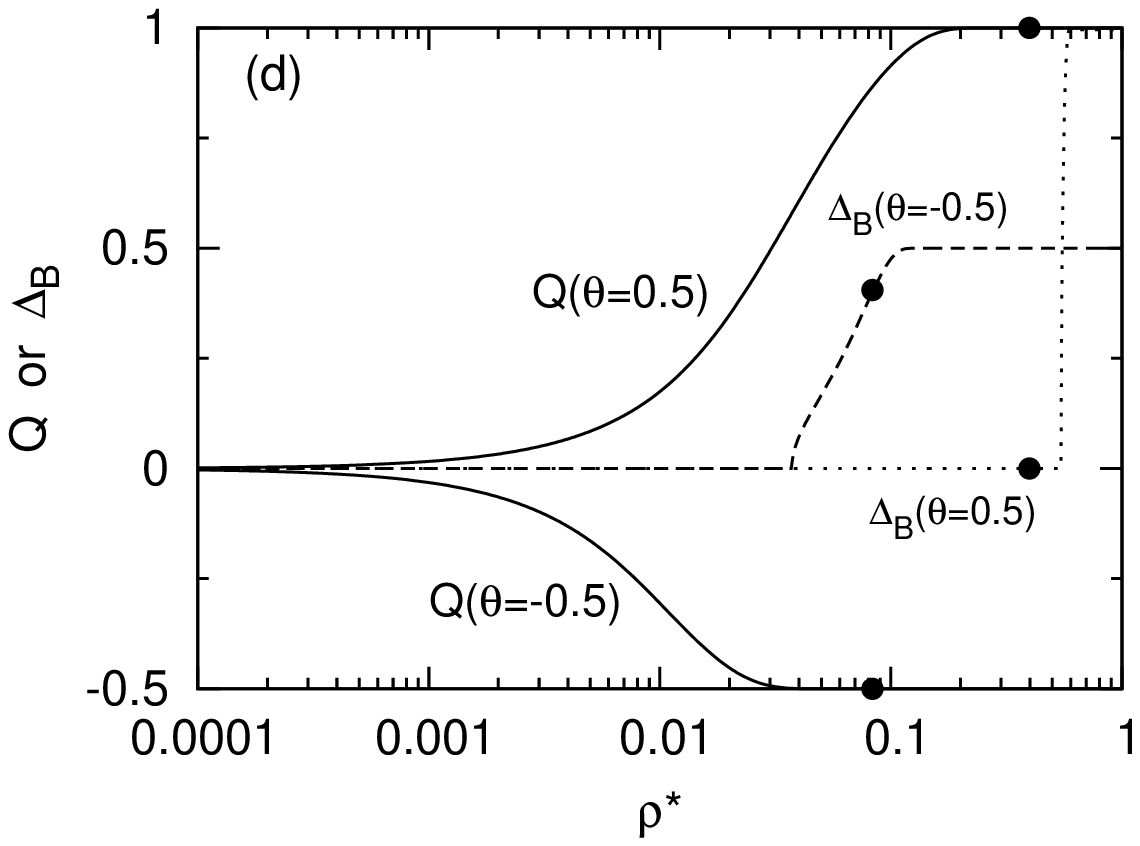}
\caption{(a) Phase diagrams in the plane scaled density $\rho^*$ vs. particle biaxiality
parameter $\theta$ for $\kappa_1=5$. Density axis is in logarithmic scale.
Solid curve: continuous N$_{\rm u}$-N$_{\rm b}$  
transition. Dashed curve: spinodal instability from the uniform to the non-uniform phases
(either K, C or S). Region of N$_{\rm b}$ stability is shaded. (b) Same as (a), but for $\kappa_1=10$. 
(c) Uniaxial $Q$ (solid curve) and biaxial $\Delta^*_Q$ (dashed curve) order parameters as a function of 
scaled density $\rho^*$ for uniaxial plate-like ($\theta=-1$) and rod-like ($\theta=1$) particles 
with $\kappa_1=10$. Density axis is in logarithmic scale. Filled circles on the curves indicate the 
instabilities to non-uniform phases. (d) Same as (c), but for plate-like ($\theta=-0.5$, dashed curve) and 
rod-like ($\theta=0.5$, dotted curve) biaxial particles.}
\label{las_cuatro}
\end{figure}

\subsection{Effect of particle biaxiality on biaxial phase}

First we chose a pair of values for the largest aspect ratio, $\kappa_1=5$ and 10, and varied particle biaxiality $\theta$ 
from $-1$ (plate-like uniaxial symmetry) to 1 (rod-like uniaxial symmetry). For each value of $\kappa_1$ and $\theta$, 
bifurcation analysis provides the values of scaled density $\rho^*$ and molar fractions 
$\{\gamma_{zx}=\gamma_{zy}\neq \gamma_{xz}=\gamma_{yz}\neq \gamma_{xy}=\gamma_{yx}\}$ 
corresponding to the N$_{\rm u}\to$ N$_{\rm b}$ bifurcation point. 
The nature (continuous vs. first order) of the transition was always checked via direct minimization 
of the free-energy density with respect to all the molar fractions $\{\gamma_{\mu\nu}\}$, which confirmed the continuous 
character of the transition from $|\gamma_{zx}-\gamma_{zy}|\sim (\rho^*-\rho^*_0)^{1/2}$ 
(with $\rho_0^*$ the scaled density at bifurcation) near and above the bifurcation point. 

The spinodal instabilities of uniform nematic phases N$_{\rm u}$, N$_{\rm b}$ with respect to density modulations of 
crystal (K), columnar (C) or smectic (S) symmetries were also obtained from the appropriate bifurcation theory 
(appendix \ref{apen3}). The results are plotted in Fig. \ref{las_cuatro}(a) 
for $\kappa_1=5$ and (b) for $\kappa_1=10$. In the first case there exists a small region (shaded in the
figure), close to 
$\theta=-1$, in which N$_{\rm b}$ is stable. The N$_{\rm u}$--N$_{\rm b}$ bifurcation line crosses the curve associated with
the spinodal instability to the non-uniform phases at $\theta\simeq -0.5$. The figure also shows the 
lack of biaxial ordering in
rod-like ($\theta>0$) particles, since the N$_{\rm b}$ bifurcation point occurs at densities higher than 
that of the C--S--K spinodal. For $\kappa_1=10$  
the region of stability of N$_{\rm b}$ is considerably enlarged (spanning the interval $\theta\alt 0$);
this is because the aspect ratio of the projected rectangles with the smallest area is larger. 
Again rod-like particles ($\theta>0$) do not exhibit biaxial ordering. 

To show the ordering properties of the system in more detail, Figs. \ref{las_cuatro}(c) and (d) contain
the behavior of the uniaxial and biaxial order parameters $Q$, $\Delta^*_Q$ and $\Delta_B$, 
as a function of $\rho^*$ for uniaxial ($\theta=\pm 1$) and biaxial ($\theta=\pm 0.5$) symmetries, 
respectively, and for $\kappa_1=10$. 
As already reported in a previous publication \cite{Yuri2}, uniaxial plates continuously become ordered
with density, their main axes lying preferentially on the surface of the
monolayer [see case $\theta=-1$ in Fig. \ref{las_cuatro}(c)]. As density increases from zero, the uniaxial 
order parameter $Q$ decreases continuously from zero 
and saturates at $-0.5$ for high densities, which means that the shortest particle axes lie on the monolayer.
In this configuration the total particle area projected on the surface is minimized 
(with a vanishingly small fraction of plates with main axes perpendicular to the monolayer). 
When $Q$ is almost saturated ($\rho^*\simeq 0.02$), the in-plane rotational symmetry of particle axes is 
broken and the system exhibits a N$_{\rm u}\rightarrow$ N$_{\rm b}$ transition.

For rod-like particles, case $\theta=1$ in Fig. \ref{las_cuatro}(c), the following behavior is observed:
the uniaxial order parameter $Q$ continuously increases from zero at vanishingly small densities, 
saturating to 1 as density increases. There are two clear differences with respect to the plate-like 
geometry: (i) axes of uniaxial particles are now preferentially oriented perpendicular to the 
monolayer (thus decreasing the total occupied area), and (ii) there is no orientational symmetry breaking:
we can discard the presence of a N$_{\rm b}$ phase for rods with $\kappa_1=10$.   

The uniaxial and biaxial order parameters are also shown as a function of $\rho^*$ for $\kappa=10$
in the case of biaxial particles with $\theta=\pm 0.5$ [see Fig. \ref{las_cuatro} (d)]. 
Here the situation is similar to the uniaxial case: 
(i) plate-like particles ($\theta=-0.5$) exhibit N$_{\rm u}$ planar ordering 
and a N$_{\rm u}$-N$_{\rm b}$ transition when $Q$ is 
almost saturated. However, the N$_{\rm b}$ phase looses its stability against non-uniform phases
at higher densities. (ii) Rod-like particles ($\theta=0.5$) possess uniaxial out-of-plane ordering 
and a direct transition from the N$_{\rm u}$ phase to the non-uniform phases (K, C or S). Note that  
N$_{\rm b}$ is metastable with respect to these phases.      

From these results we can conclude that, contrary to intuition, the main effect of particle biaxiality 
in plate or rod monolayers is the destabilization of the N$_{\rm b}$ phase: note in
Figs. \ref{las_cuatro}(a) and (b) how the shaded region of N$_{\rm b}$ stability, bounded by the two 
spinodal curves, shrinks as $\theta$
increases from the uniaxial case $\theta=-1$. There is a clear physical interpretation of this 
behavior. For uniaxial plates, with dimensions $\sigma\times\sigma\times L$ ($L<\sigma$), 
there are two identical rectangular and mutually perpendicular projections of dimensions $L\times\sigma$,
which have different molar fractions for a given density, and consequently a N$_{\rm b}$ appears. 
The other (large) projection, of dimensions $\sigma\times\sigma$, has a vanishingly small molar fraction. 
When particle biaxiality increases keeping fixed the largest aspect ratio ($\kappa_1=\sigma_1/\sigma_3=10$; 
without loss of generality we suppose $\sigma_3$ to be constant), decreasing 
$\sigma_2$ from $\sigma_1$, the original projected rectangular species of equal areas becomes now different, 
with dimensions $\sigma_1\times\sigma_3$ (intermediate species) and $\sigma_2\times\sigma_3$ (smallest species). 
Note that biggest species, that with dimensions $\sigma_1\times\sigma_2$, will continue to have a vanishingly 
small molar fraction. To minimize the excluded volume interactions between particles, the fraction of 
$\sigma_2\times\sigma_3$ species should increase with respect to the other, and the total density has
to increase to stabilize the N$_{\rm b}$ phase (we remind that a larger aspect ratio favours the
N$_{\rm u}$-N$_{\rm b}$  symmetry breaking).

\begin{figure}
\includegraphics[width=5.2cm]{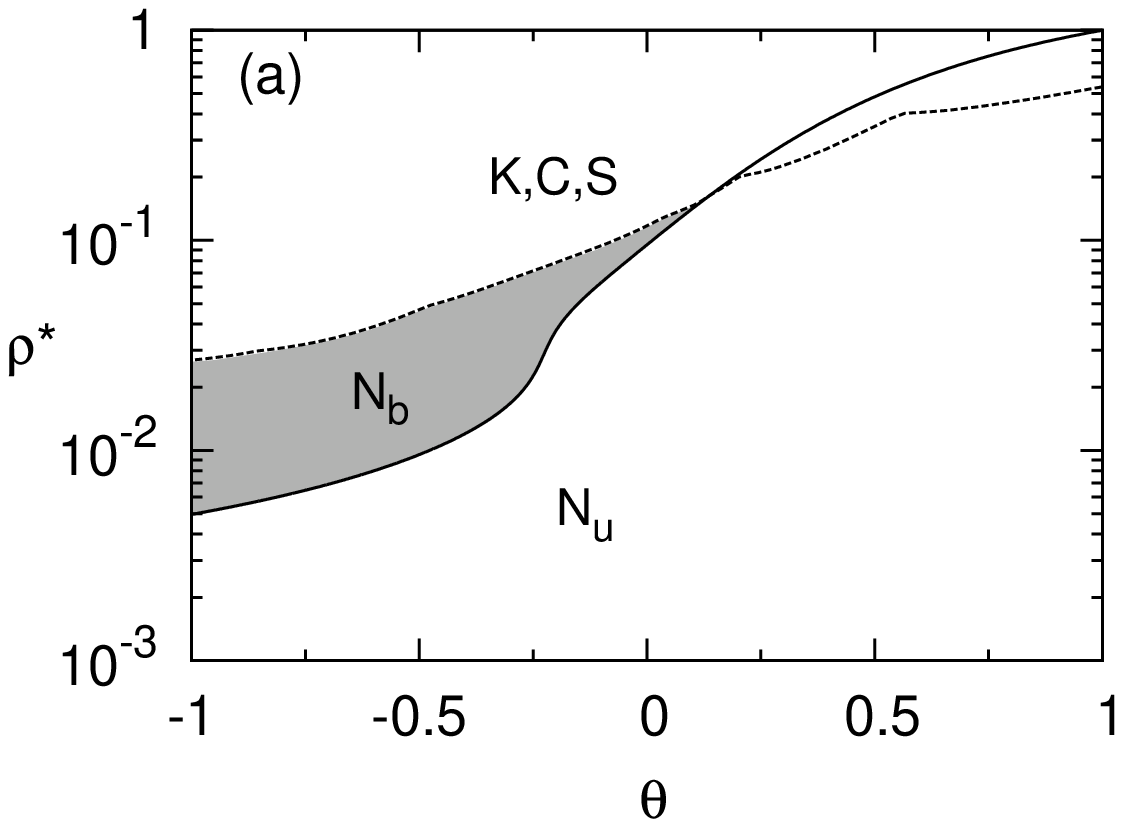}
\includegraphics[width=5.2cm]{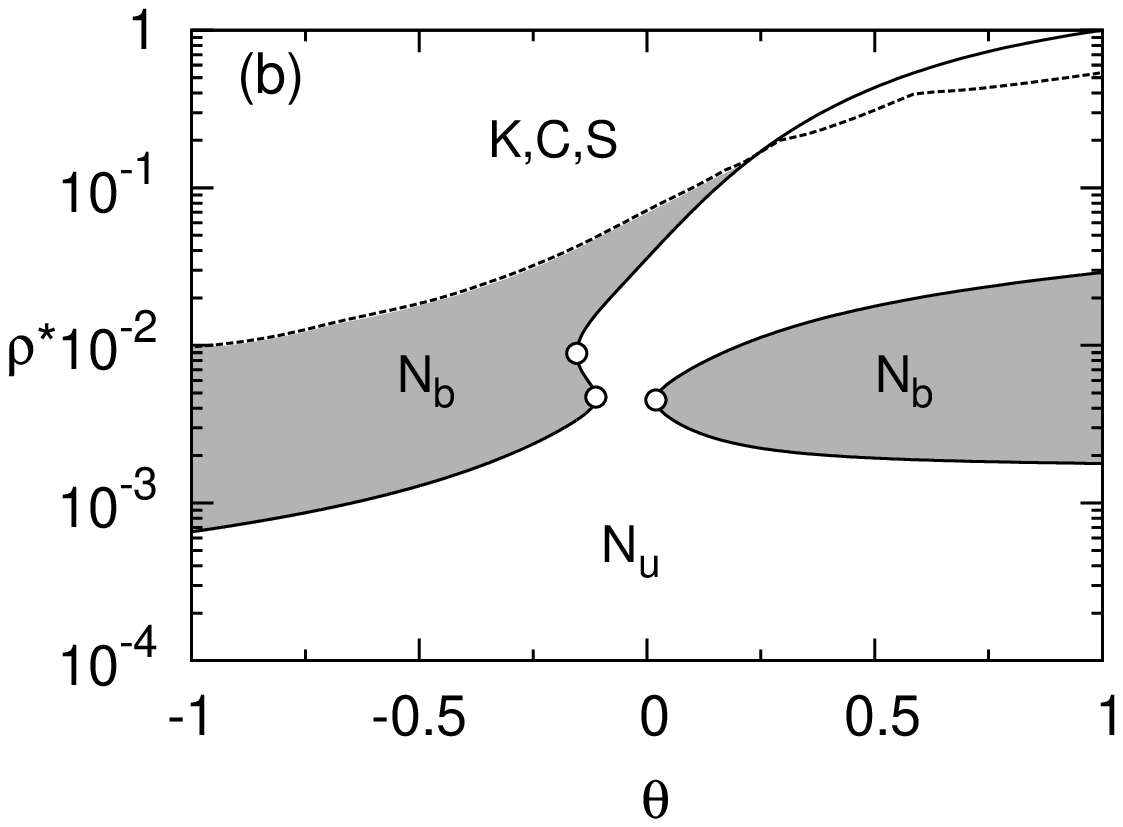}
\includegraphics[width=5.2cm]{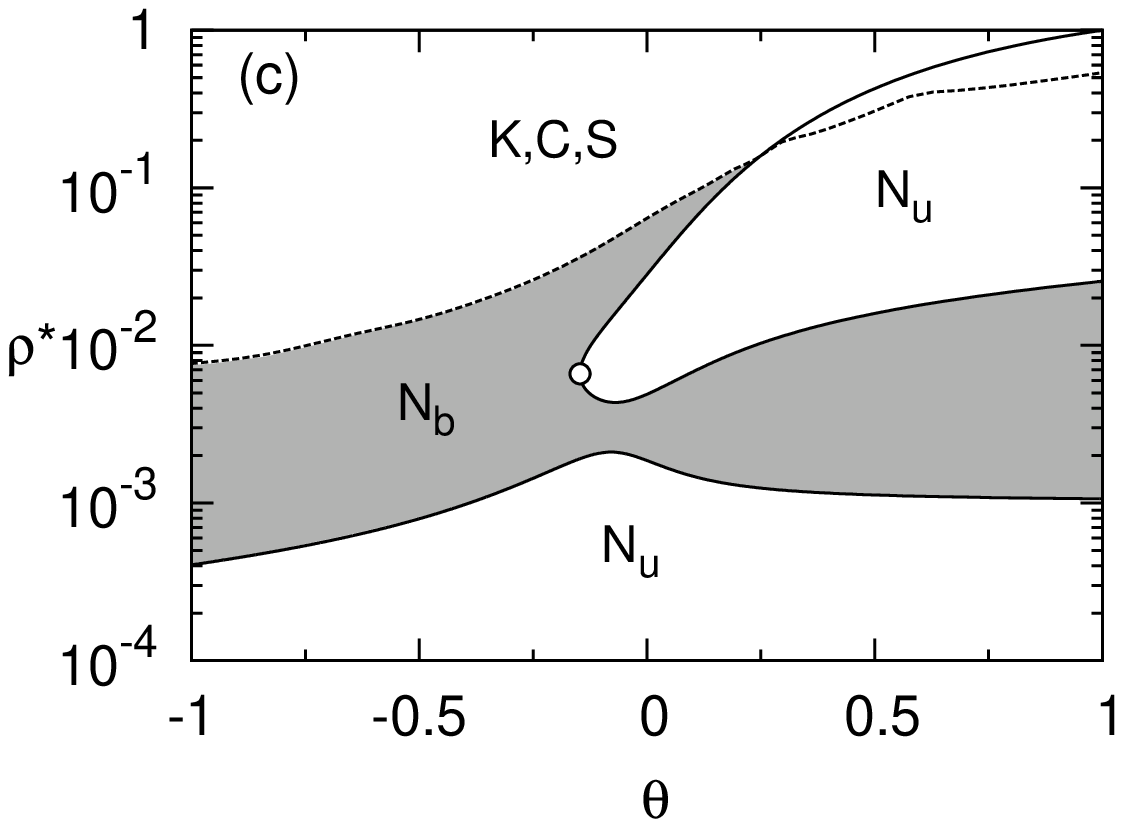}
\includegraphics[width=5.2cm]{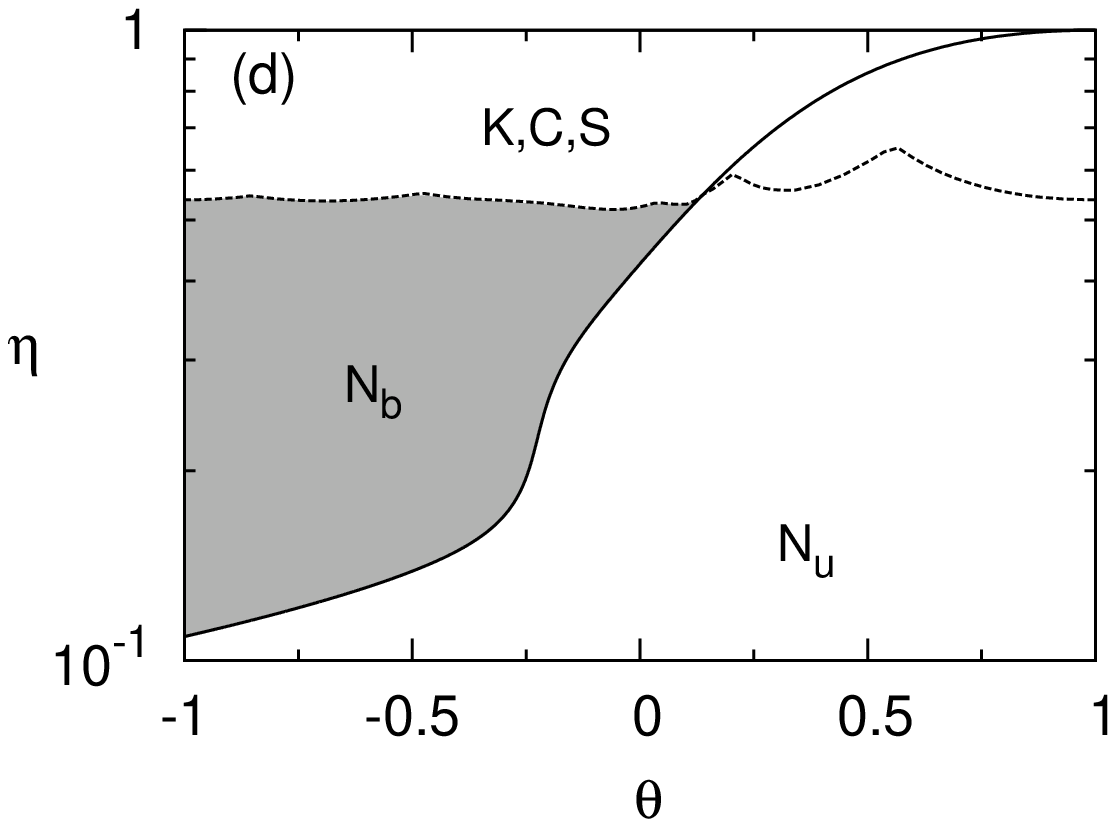}
\includegraphics[width=5.2cm]{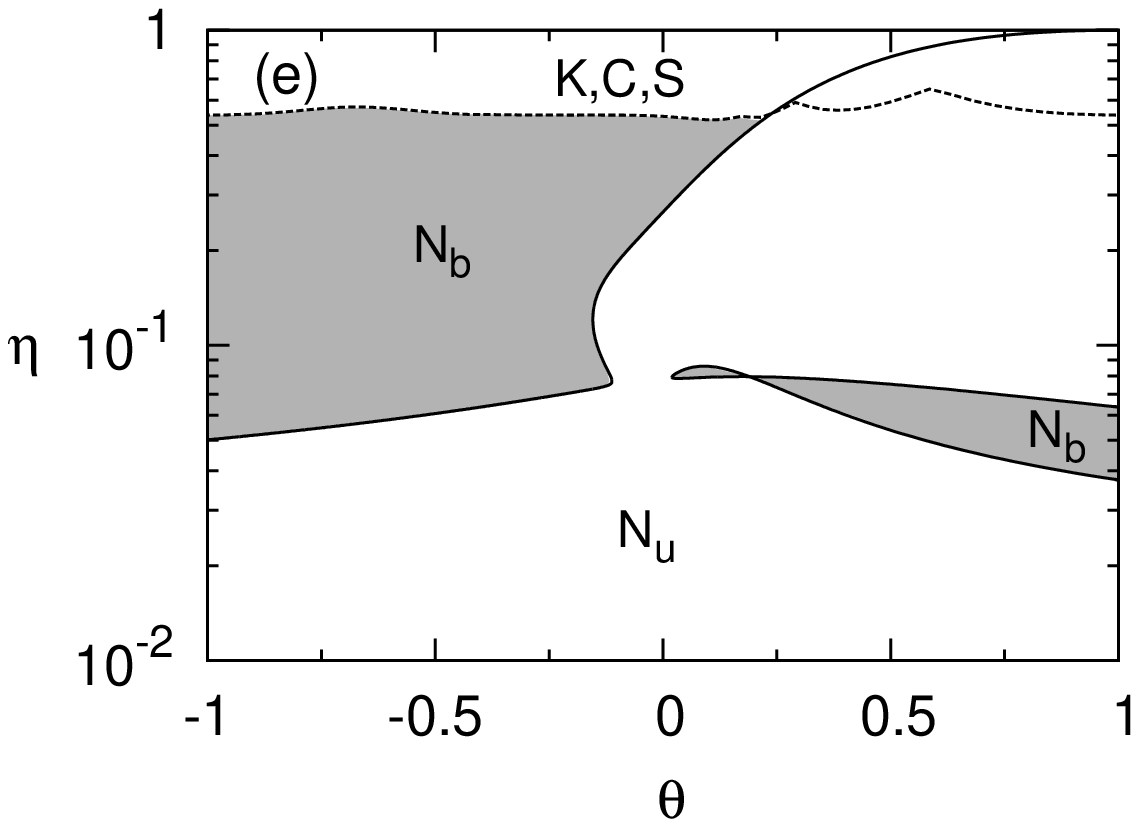}
\includegraphics[width=5.2cm]{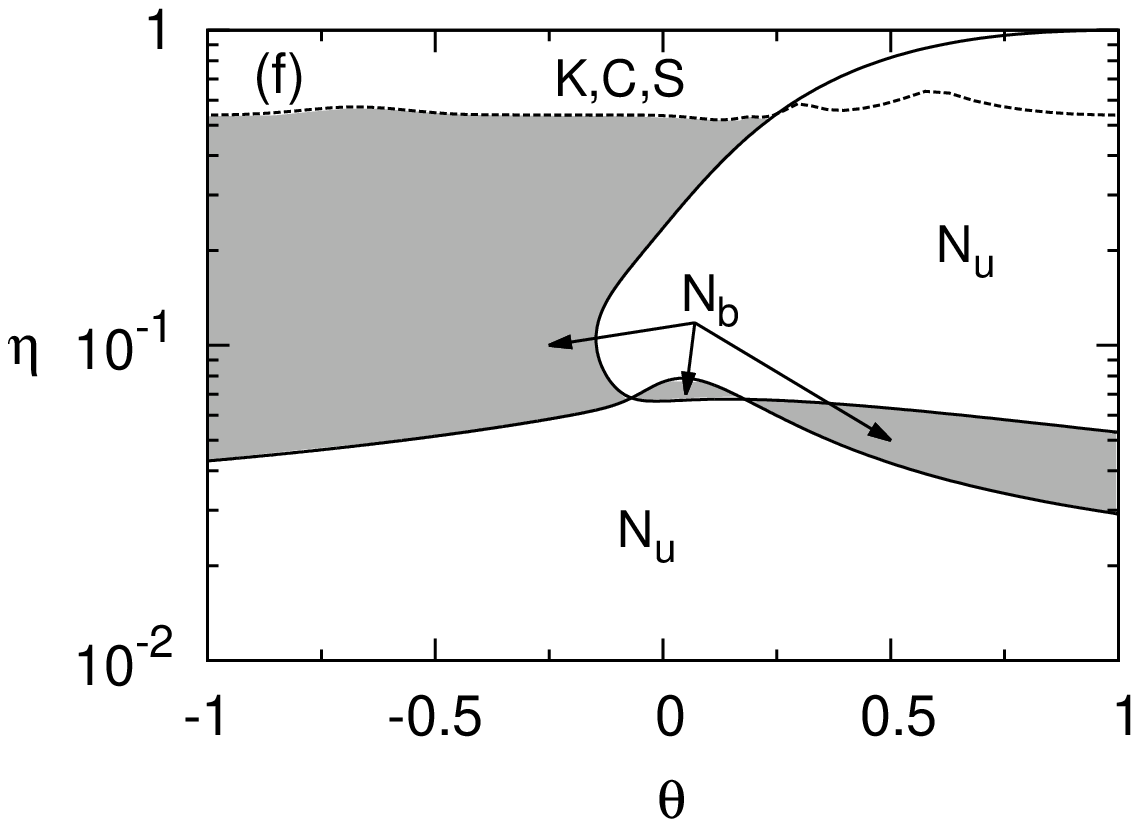}
\caption{Phase diagrams in the $\rho^*-\theta$ [(a), (b), and (c)] and $\eta-\theta$ [(d), (e), and (f)] planes 
for $\kappa_1=20$ [(a) and (d)], 55 [(b) and (e)] and 70 [(c) and (f)]. The regions of stability of different 
phases are correspondingly labelled. Regions of stability of the N$_{\rm b}$ phase are shaded.
Curves have the same meanings as in Fig. \ref{las_cuatro} (a) and (b). 
Open circles over the curves represent the positions of critical end-points.}
\label{las_6}
\end{figure}

It is fruitful to compare (at least qualitatively) our results with those of the recent experiment of goethite nanorods confined 
between the bilayers of a lamellar phase made from nonionic surfactant \cite{30,31,32}. These particles 
orient perpendicular to an applied magnetic field along the 
lamellae axis so that negative uniaxial order parameters can be obtained, resulting in stacked sheets of
liquid-like quasi-two-dimensional rods. Particle sizes
were estimated by optical and X-ray diffraction methods to be $315\times 38\times 18$ nm$^3$ resulting, 
in our notation, 
in aspect ratios $\kappa_1=17.5$ and $\kappa_2=2.1$, and $\theta=0.37$ (i.e. relatively biaxial
particle sizes). Rod interactions are approximately
hard, but interact with the lamellae in complex ways, probably resulting in effective attractions
between the rods in a sheet; intersheet interactions also exist, although they are probably weak. The 
authors find an `isotropic' phase (corresponding to the uniaxial nematic phase N$_{\rm u}$ in our
monolayer) and a `nematic' phase (our biaxial N$_{\rm b}$ phase) and suggest a possible continuous
phase transition between the two at a packing fraction which was not possible to estimate in the
experiment.  
This particle geometry would correspond closely to the phase diagram of Figs. \ref{las_6}(a) and (d).
In our diagram, the experimental value of $\theta$ is slightly larger than the predicted 
limiting point for the biaxial phase. A number of factors could explain the difference: 
modified attractive interactions and size polydispersity in the experimental nanorod system, both
of which could enhance the stability of the biaxial phase, and/or defects in the theoretical approach.

\subsection{Topology of phase diagram}

In this section the topology of the phase diagram as a function of $\kappa_1$ is analysed. 
Figs. \ref{las_6} show phase diagrams in the $\rho^*-\theta$ [(a--c)] and $\eta-\theta$ [(d--f)]
planes for $\kappa_1=20$, 55 and 70. For $\kappa_1=20$ [(a) and (d)] the phase diagrams retain 
the same topology as for the case $\kappa_1=10$, see Fig. \ref{las_cuatro}(b). 

As shown in our recent work \cite{Yuri2}, a monolayer of uniaxial rods in the restricted-orientation
approximation exhibits a peculiar phase behaviour for $\kappa_1=21.34$, with 
a reentrant N$_{\rm u}$ phase and an intermediate N$_{\rm b}$ phase. This behaviour persists in the
case of biaxial rods. For example, Figs. \ref{las_6}(b) and (e) pertain to the case $\kappa_1=55$,
and the aforementioned system would be similar to the case $\theta=1$ (uniaxial rods).
The N$_{\rm b}$ stability region shrinks for increasing 
particle biaxiality (decreasing $\theta$), totally disappearing at a critical-end point 
(shown with open circle). The presence of a biaxial phase in monolayers of uniaxial rods is easy to explain:
For high aspect ratios and densities such that the total packing fraction of the projected rectangular species 
$L\times\sigma$ is close to $\eta_{\rm 2D}$ (that of the I-N transition 
of hard rectangles in 2D), an orientational symmetry breaking at the surface of the monolayer takes places. 
Of course the presence of the 
square, $\sigma\times\sigma$, species should be taken into account. However, at low densities 
and high aspect ratios, the packing fraction of squares is small compared to that of rectangles. 
When the total density is increased,
the packing fraction of squares increases (as uniaxial nematic ordering is promoted), 
while the packing fraction of rectangles decreases. Then the packing fraction of rectangles 
jumps below $\eta_{\rm 2D}$, and consequently the N$_{\rm b}$ phase looses its stability with 
respect to the N$_{\rm u}$ phase. 

Now we discuss the stability region of the biaxial nematic phase on the prolate side. 
When particle biaxiality is increased ($\theta$ decreases from 1),
rectangular species becomes inequivalent and the largest one, of dimensions $\sigma_1\times\sigma_2$, 
rapidly decreases in molar fraction 
with respect to the intermediate one, of dimensions $\sigma_1\times\sigma_3$.
Therefore the total density should increase so that the total area fraction of the projected rectangular 
becomes on the order of $\eta_{\rm 2D}$, and the 
N$_{\rm u}$-N$_{\rm b}$ transition density increases. On the other hand, the alignment of particles along $z$ 
is enhanced with the increased biaxiality such that the packing fraction of the 
smallest species grows at the expense of the other two, and consequently the N$_{\rm b}$-N$_{\rm u}$ 
transition curve moves to lower densities. For a particular value of particle biaxiality the two
transition curves, for the N$_{\rm u}$-N$_{\rm b}$ and N$_{\rm b}$-N$_{\rm u}$ transitions, coalesce into
a single critical end-point [see Fig. \ref{las_6}(b)]. 

\begin{figure}
\includegraphics[width=8cm]{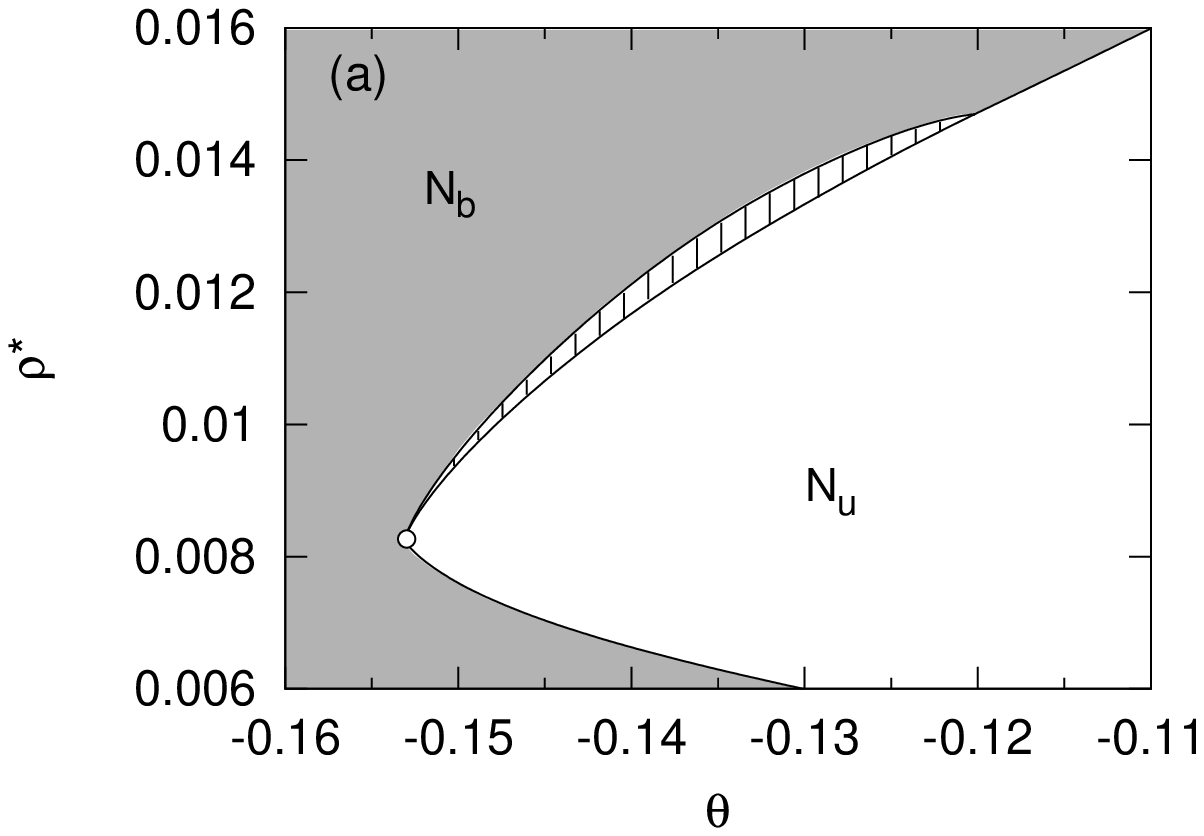}
\includegraphics[width=8cm]{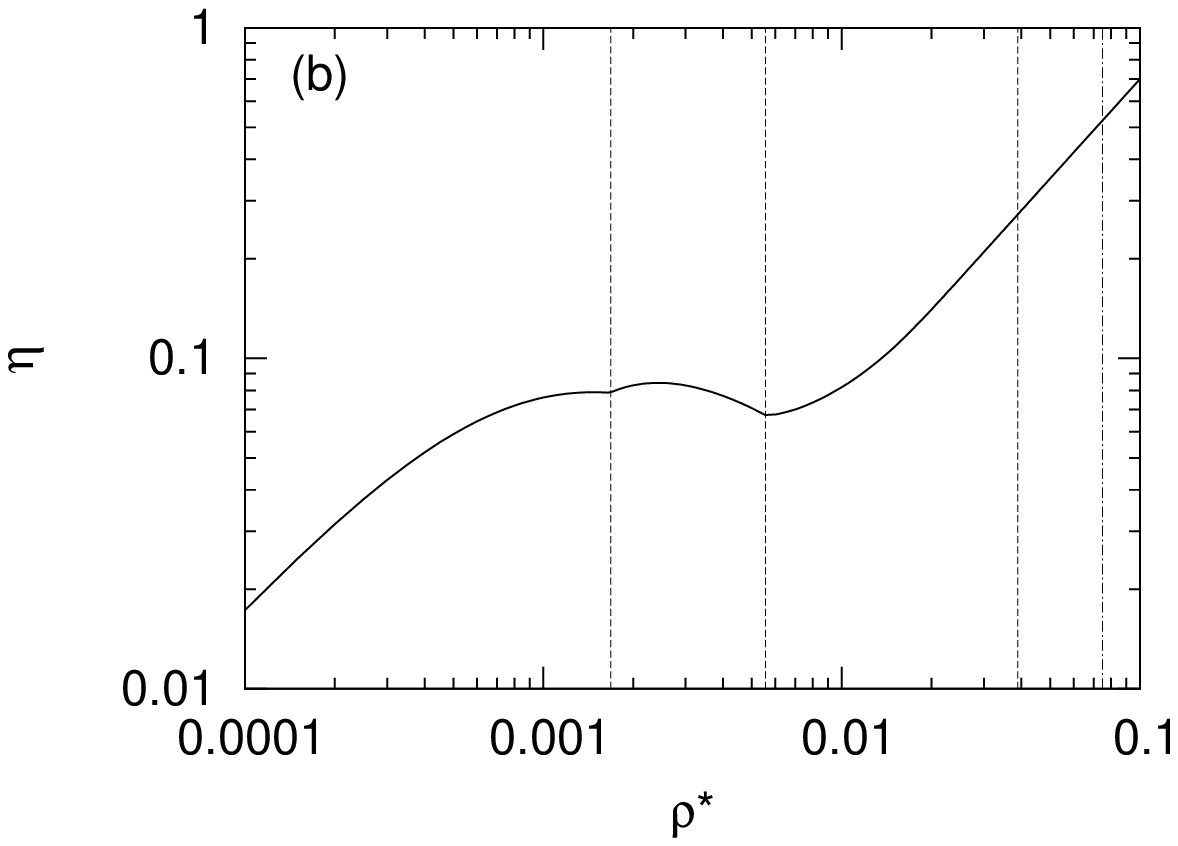}
\caption{(a) Detail of Fig. \ref{las_6}(b) in which a region  
of first-order N$_{\rm u}$--N$_{\rm b}$ transitions (hatched area) is observed. (b) $\eta$ vs. $\rho^*$ 
for $\kappa_1=70$ and $\theta=0.04$ [see Fig. \ref{las_6}(c)] showing the packing fraction inversion. 
Vertical lines show the values of $\rho^*$ corresponding to phase transitions between different phases.
}
\label{detail}
\end{figure}

It is interesting to note from Figs. \ref{las_6}(a--c) that the N$_{\rm u}$--N$_{\rm b}$ 
spinodal in the region $\theta<0$ deforms as $\kappa_1$ increases and eventually develops
a loop at $\theta\simeq 0$ [see Fig. \ref{las_6}(b) for
$\kappa_1=55$]. This means that there is a small $\theta$-interval 
(between the two critical end-points shown with open circles) where both nematic phases are reentrant.
Also, there is a particular value of aspect ratio, $\kappa_1^*\simeq 60$, for which the two critical 
end-points in the plate-like ($\theta<0$) and rod-like ($\theta>0$), on the corresponding
N$_{\rm b}$ spinodals, coalesce into a single 
point. For $\kappa_1>\kappa_1^*$ a density gap appears in the phase diagram where 
the N$_{\rm b}$ is stable for any $\theta$, as shown in Fig. \ref{las_6}(c) for $\kappa_1=70$. 
Now both nematic phases are reentrant in wide intervals of $\theta$.
  
We should mention that the nature of the N$_{\rm u}$-N$_{\rm b}$ transition is always continuous, 
except for $\kappa_1> 40$ in a very small range of particle biaxiality corresponding to the density loop 
mentioned above. This is shown in Fig. \ref{detail}(a) 
where a detail of the phase diagram for $\kappa_1=55$ is shown. The hatched area represents 
the N$_{\rm u}$-N$_{\rm b}$ coexistence region. 

For high enough $\kappa_1$ and particles with $\theta\simeq 0$ the phase diagrams present an 
interesting feature, namely a packing-fraction inversion. This is shown  
in Fig. \ref{las_6}(e) and (f) for $\kappa_1=55$ and $70$.
In this region the lower N$_{\rm u}$ $\to$ N$_{\rm b}$ and upper 
N$_{\rm b}$ $\to$ N$_{\rm u}$ transition curves in the $\rho^*-\theta$ plane change 
their relative locations when plotted 
in the $\eta-\theta$ plane. This peculiar phenomenon can be clearly visualized in 
Fig. \ref{detail}(b), where the packing fraction $\eta$ is plotted against $\rho^*$. 
As we can see, once the N$_{\rm u}$-N$_{\rm b}$ transition takes place, the transition  
packing fraction exhibits a maximum and then decreases down to the value at the 
N$_{\rm b}$--N$_{\rm u}$ transition. For larger $\rho^*$ the packing fraction exhibits the usual 
monotonic behaviour. This effect can be explained by resorting to Eqn. 
(\ref{la_packing}), which shows that $\eta$ 
is a function of $\rho^*$ and the two order parameters $Q$ and $B$. It is then possible for the
packing fraction to decrease with $\rho^*$ when the order parameters
are positive and increase sufficiently strongly with $\rho^*$ 
(i.e. uniaxial ordering is strongly promoted so that the number of particles of the species with 
the smallest projected area increases rapidly enough), in such a way that  
the total increase in the number of particles is compensated.

\begin{figure}
\includegraphics[width=8cm]{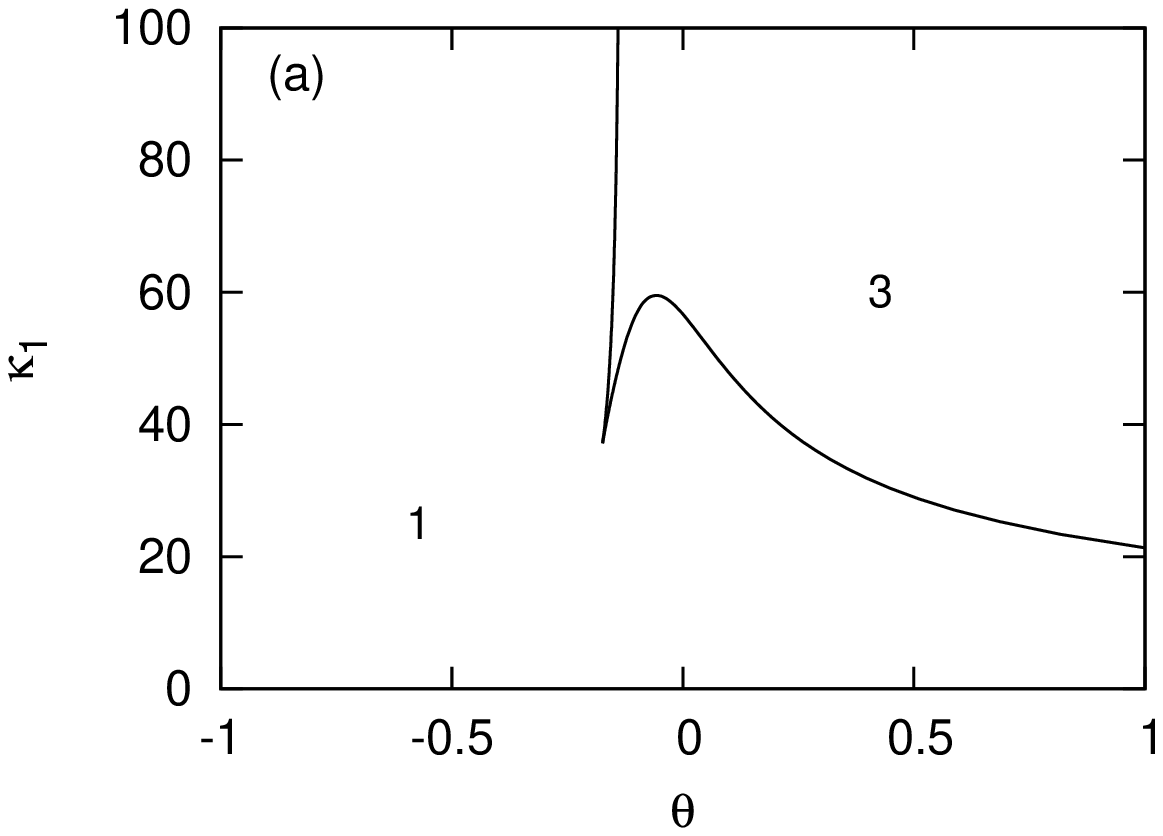}
\includegraphics[width=8cm]{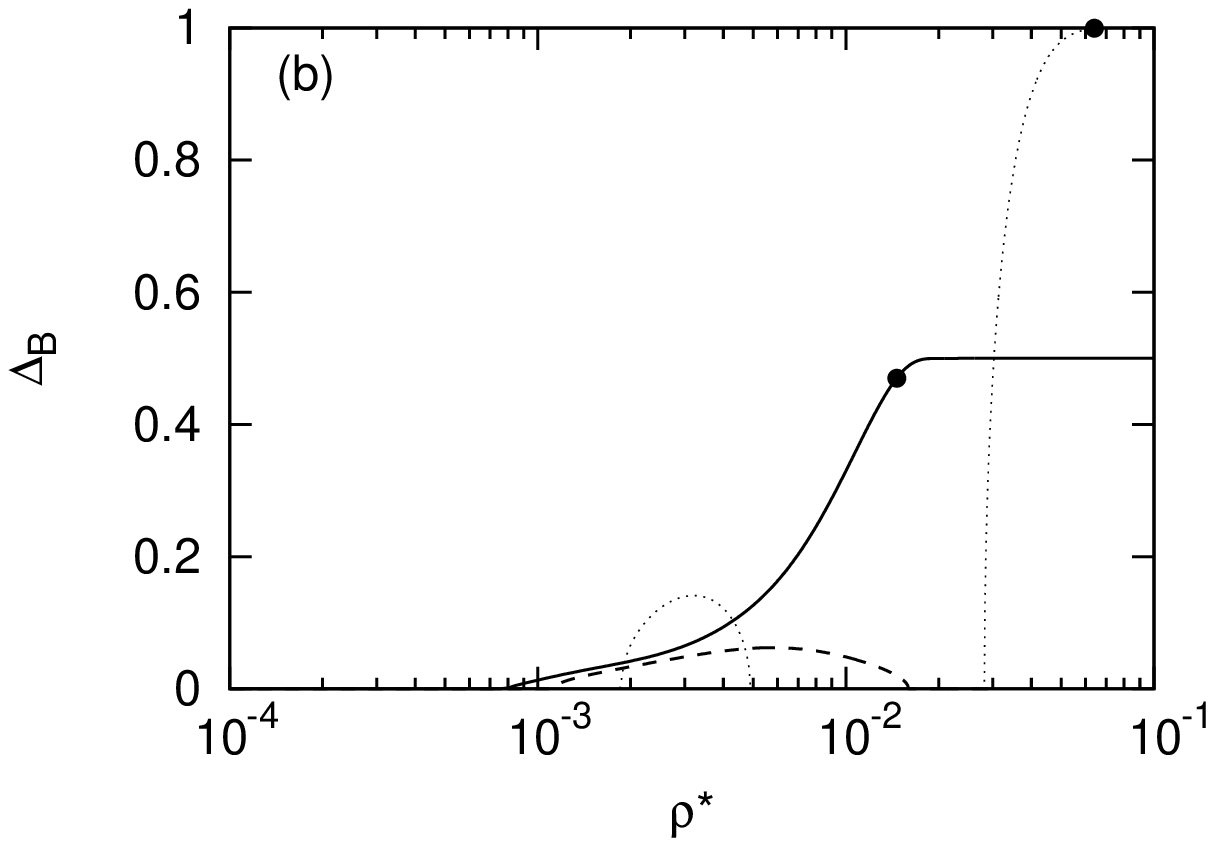}
\caption{(a) Location of critical end-points in the $\kappa_1-\theta$ plane. 
The curve separates regions where one or three phase transitions between uniform phases take place. 
(b) Biaxial order parameter $\Delta_B$ vs. $\rho^*$ for $\theta=-0.5$ (solid), 
0 (dotted) and 0.5 (dashed) for $\kappa_1=70$. Filled circles
on the curves indicate the instabilities to non-uniform phases.}
\label{son_dos}
\end{figure}

One interesting feature of the phase diagrams shown in Fig. \ref{las_6} is that, for particular
values of the parameters $(\kappa_1,\theta)$,  
the total number of transitions between uniform phases (N$_{\rm u}$ and N$_{\rm b}$) can be one or 
three (the latter case associated with reentrant phases). 
The curve in the $\kappa_1-\theta$ plane separating both regions is just the continuous boundary of critical 
end-points (see appendix \ref{apen2} for details on their calculation), and is
plotted in Fig. \ref{son_dos}(a), where the regions corresponding to one or three phase transitions are 
correspondingly labelled.

To finish this section, we compare in Fig. \ref{son_dos}(b) the biaxial orientational order, as
measured by the biaxial order parameter $\Delta_B$, of plate-like ($\theta=-0.5$, solid curve), 
perfectly biaxial ($\theta=0$, dotted curves) and rod-like 
($\theta=0.5$, dashed curve) particles, all of them having $\kappa_1=70$. 
There are important differences between the three cases: 
while the biaxial order of plate-like particles increases from the bifurcation point and 
finally saturates at its maximum value, 
rod-like particles exhibit a rather small biaxial order in the range of 
densities where the N$_{\rm b}$ phase is stable. Finally, for $\theta=0$ 
a small region of biaxial order (with 
relatively small order parameter) is followed by a second transition to a second N$_{\rm b}$ phase
possessing a high degree of biaxiality up to the transition to a non-uniform phase. This trend is general
for any $\kappa_1>21.34$: The N$_{\rm b}$ phase of rod-like particles exhibits only a small degree of 
global biaxial ordering, quantified through the order parameter $\Delta_B$. Therefore, its stability is
questionable for the freely-rotating case.

\section{Conclusions}
\label{IV}

In this work we studied the effect 
of particle biaxiality on the phase behaviour of liquid-crystal colloidal monolayers, using a 
fundamental-measure density-functional theory for hard board-like biaxial particles
with restricted orientations. Various phase diagrams were obtained for different values of the 
two parameters that describe the particle shape: the largest particle aspect ratio $\kappa_1$ 
and the particle biaxiality $\theta$. This study is an extension of our previous work in which 
monolayers of uniaxial rod-like and plate-like particles were analysed \cite{Yuri2}. 
Contrary to expected, particle biaxiality 
destabilizes the biaxial phase in the cases where the latter is present, 
a phenomenon directly related with the competition between 
(i) the biaxiality promoted by the two-dimensional spatial constraint
on particle centres of mass,  and (ii) the biaxial ordering promoted by particle biaxiality for high
enough densities.
For biaxial particles the rectangular projected areas are inequivalent, and the mixing entropy 
stabilizes, mainly for plate-like geometry, the 2D isotropic phase. 

For rod-like geometry the N$_{\rm b}$ phase has 
a small degree of biaxial orderer and occurs in a narrow interval of densities. Again an increase in
particle biaxiality reduces the stability interval which eventually disappears at a critical end point. 
For high enough values of the largest aspect ratio, $\kappa_1\simeq 60$, the phase diagram exhibits 
a density gap in which the N$_{\rm b}$ is stable for any value of particle biaxiality $\theta$. 
The transitions between nematic phases are continuous, except for a small 
range of values of $\theta$ about zero and for large values of $\kappa_1$, where a first-order 
N$_{\rm u}$-N$_{\rm b}$ transition appears. A packing 
fraction inversion phenomenon also exists. The rapid increase of particle alignment along $z$, 
resulting in a large fraction of the projected species with the smallest area, compensates the total 
increment in number of particles, resulting in a decrease of $\eta$ with $\rho^*$.  

The presence of a N$_{\rm b}$ phase in the rod-like region of the phase diagram ($\theta>0$) 
should be taken with care, as it could be a direct consequence of the restriction on particle orientations. 
As shown by Monte Carlo simulations and Parsons-Lee density-functional theory, uniaxial freely-rotating
plate-like ellipsoidal particles adsorbed on a monolayer without orientational restrictions
do exhibit a N$_{\rm b}$ phase, while their rod-like counterparts do not \cite{unpublished}. 
However it would be necessary to explore a larger variety of particle geometries, without imposing 
orientational constraints, to finally discard the presence of a biaxial nematic phase.     

We hope that our study will serve as a guide for future experimental studies of confined board-shaped 
colloidal systems, such as goethite nanorods \cite{30,31,32} and the recently synthesized lead carbonate 
nanoplatelets \cite{33}. 

\acknowledgments
We gratefully acknowledge illuminating discussions with G. Odriozola.
Financial support from MINECO (Spain) under grants FIS2010-22047-C01 and FIS2010-22047-C04 is acknowledged.
SV acknowledges the financial support of the Hungarian State and the European Union under the 
TAMOP-4.2.2.A-11/1/KONV-2012-0071.

\appendix
\section{Density-functional Theory}
\label{apen1}

We use a density functional (DF) based on fundamental-measure theory for the Zwanzig model 
(particle orientations along the three Cartesian axes) which fulfills the 3D$\to$2D dimensional crossover 
\cite{Cuesta}. In this formalism the excess part of free-energy density depends on a set of weighted 
densities calculated by convoluting the density profiles of the two-dimensional projections of the six species 
with certain weighting functions, the latter depending on the geometry of a single particle: 
\begin{eqnarray}
&&n_{\alpha}({\bm r})=\sum_{\mu\nu}\int d{\bm r}'\rho_{\mu\nu}({\bm r}')
\omega^{(\alpha)}_{\mu\nu}({\bm r}-{\bm r}'),\nonumber\\
&&\omega^{(0)}_{\mu\nu}({\bm r})=\frac{1}{4}\delta\left(\frac{\sigma^x_{\mu\nu}}{2}-|x|\right)
\delta\left(\frac{\sigma^y_{\mu\nu}}{2}-|y|\right),\nonumber\\
&&\omega^{(1x)}_{\mu\nu}({\bm r})=\frac{1}{2}\Theta\left(\frac{\sigma^x_{\mu\nu}}{2}-|x|\right)
\delta\left(\frac{\sigma^y_{\mu\nu}}{2}-|y|\right),\nonumber\\
&&\omega^{(1y)}_{\mu\nu}({\bm r})=\frac{1}{2}\delta\left(\frac{\sigma^x_{\mu\nu}}{2}-|x|\right)
\Theta\left(\frac{\sigma^y_{\mu\nu}}{2}-|y|\right),\nonumber\\
&&\omega^{(2)}_{\mu\nu}({\bm r})=\Theta\left(\frac{\sigma^x_{\mu\nu}}{2}-|x|\right)
\Theta\left(\frac{\sigma^y_{\mu\nu}}{2}-|y|\right),
\end{eqnarray}
where $\delta(x)$ and $\Theta(x)$ are the Dirac delta and Heaviside functions, respectively, while we have introduced the 
tensor $\sigma_{\mu\nu}^{\tau}=\sigma_3+(\sigma_1-\sigma_3)\delta_{\tau\mu}+(\sigma_2-\sigma_3)\delta_{\tau\nu}$ 
(with $\tau=x,y$ and $\delta_{\tau\mu}$ the Kronecker delta). In the uniform limit we obtain 
$n_{\alpha}=\sum_{\mu\nu}\rho_{\mu\nu}{\cal M}_{\mu\nu}^{(\alpha)}$, with ${\cal M}_{\mu\nu}^{(\alpha)}=\int d{\bm r} 
\omega_{\mu\nu}^{(\alpha)}({\bm r})$ 
the fundamental measures of the 2D particle projections:
\begin{eqnarray}
{\cal M}^{(0)}_{\mu\nu}=1,\quad {\cal M}_{\mu\nu}^{(1\tau)}=\sigma^{\tau}_{\mu\nu},\quad {\cal M}_{\mu\nu}^{(2)}=\sigma^x_{\mu\nu}
\sigma_{\mu\nu}^y.
\end{eqnarray}
The excess part of the scaled free-energy density for a 2D mixture of six particle projections,  
\begin{eqnarray}
\Phi_{\rm exc}^*\equiv \beta {\cal F}_{\rm exc}\sigma_3^2/A=\sigma_3^2\left(-n_0\ln(1-n_2)+\frac{n_{1x}n_{1y}}{1-n_2}\right),
\end{eqnarray}
(with ${\cal F}_{\rm exc}$ the uniform limit of the excess part of the DF and $A$ the total area) can be written as 
\begin{eqnarray}
\Phi_{\rm exc}^*=\rho^*\left[-\ln(1-\eta)+y\Psi_{1x}\Psi_{1y}\right],
\label{exceso}
\end{eqnarray}
where the scaled density is defined as $\rho^*=\rho\sigma_3^2$ and the packing fraction, $\eta=\rho^* \Psi_2$, 
is the uniform limit of the weighted density $n_2({\bm r})$. Also we have defined $y=\rho^*/(1-\eta)$, and the following functions  
\begin{eqnarray}
&&\Psi_{1x}=\left(\gamma_{xy}+\gamma_{xz}\right)\kappa_1+
\left(\gamma_{yx}+\gamma_{zx}\right)\kappa_2+\gamma_{zy}+\gamma_{yz},\\
&&\Psi_{1y}=\left(\gamma_{yx}+\gamma_{yz}\right)\kappa_1+
\left(\gamma_{xy}+\gamma_{zy}\right)\kappa_2+\gamma_{zx}+\gamma_{xz},\\
&&\Psi_2=\left(\gamma_{xy}+\gamma_{yx}\right)\kappa_1\kappa_2+
\left(\gamma_{xz}+\gamma_{yz}\right)\kappa_1+\left(\gamma_{zx}+\gamma_{zy}\right)\kappa_2.
\end{eqnarray}
The ideal part of the free-energy density in reduced units is 
\begin{eqnarray}
\Phi_{\rm id}^*\equiv \beta {\cal F}_{\rm id}\sigma_3^2/A=
\rho^*\left[\ln \rho^*-1+\sum_{\mu,\nu}\gamma_{\mu\nu}\ln\gamma_{\mu\nu}\right].
\end{eqnarray}
The minimization of the total free-energy density, $\Phi^*=\Phi_{\rm id}^*+\Phi_{\rm exc}^*$, with respect to the molar 
fractions $\gamma_{\mu\nu}$, together 
with the constraint $\sum_{\mu,\nu}\gamma_{\mu\nu}=1$, provide the following set of equations 
that have to be solved to obtain their equilibrium values:
\begin{eqnarray}
&&\gamma_{\mu\nu}=\frac{e^{-\chi_{\mu\nu}}}{\sum_{\alpha\beta} e^{-\chi_{\alpha\beta}}} 
\label{itera},\\
&&\chi_{\mu\nu}=
y\left[\Psi_{1x}\kappa_{\mu\nu}^y+\Psi_{1y}\kappa_{\mu\nu}^x+
\left(1+y\Psi_{1x}\Psi_{1y}\right)\kappa_{\mu\nu}^x\kappa_{\mu\nu}^y\right],
\end{eqnarray}
where we have denoted $\kappa_{\mu\nu}^{\tau}=1+(\kappa_1-1)\delta_{\tau\mu}+(\kappa_2-1)\delta_{\tau\nu}$.  

The chemical potentials of the species $\tau\nu$ evaluated at the equilibrium $\{\gamma_{\mu\nu}^{(\rm eq)}\}$ are 
\begin{eqnarray}
\beta \mu_{\tau\nu}=\beta \mu_0=\ln\left(\frac{y}{\sum_{\alpha\beta} e^{-\chi_{\alpha\beta}}}\right),\quad \forall \ \tau,\nu
\end{eqnarray}
Finally, the pressure in reduced units can be computed as
\begin{eqnarray}
p^*\equiv\beta p \sigma_3^2=y+y^2\Psi_{1x}\Psi_{1y}.
\end{eqnarray}
Both quantities are required to calculate the coexistence densities in case of first-order phase transitions. 

\section{Bifurcation to the biaxial phase}
\label{apen2}

Here we perform a bifurcation analysis from the uniaxial nematic (N$_{\rm u}$) to 
the biaxial nematic (N$_{\rm b}$) phase. The latter phase has two nematic directors, perpendicular 
and parallel to the monolayer, respectively. By solving Eqs. (\ref{set}) and (\ref{packing}) below
we find the values of the scaled density 
$\rho^*$ and two independent molar fractions at the bifurcation (spinodal). Note that in case of continuous 
N$_{\rm u}$-N$_{\rm b}$ phase transitions, this formalism provide the exact location of the transition point.   

Let us define the new variables $u_{\pm}=(\gamma_{zx}\pm\gamma_{zy})/2$, $v_{\pm}=(\gamma_{xz}\pm\gamma_{yz})/2$, and 
$r_{\pm}=\left(\gamma_{xy}\pm\gamma_{yx}\right)/2$  which for N$_{\rm u}$ symmetry ($\gamma_{zx}=\gamma_{zy}$, 
$\gamma_{xz}=\gamma_{yz}$ and $\gamma_{xy}=\gamma_{yx}$)
are equal to $\gamma_{zx}$, $\gamma_{xz}$ and $\gamma_{xy}$ for the $(+)$ sign, and strictly zero 
for the $(-)$ sign. Also let us define the quantities
\begin{eqnarray}
s_{\pm}=u_{\pm}(\kappa_2\pm 1)+v_{\pm}(\kappa_1\pm 1)+r_{\pm}(\kappa_1\pm\kappa_2).
\end{eqnarray}
Then we find that
\begin{eqnarray}
\Psi_{1x}\Psi_{1y}=s_+^2-s_-^2,\quad \Psi_2=2\left(u_+\kappa_2+v_+\kappa_1+r_+\kappa_1\kappa_2\right).
\end{eqnarray}
The ideal part of the free-energy density, in these new variables, has the form 
\begin{eqnarray}
&&\Phi_{\rm id}^*=\rho^*\left\{\ln\rho^*-1+\sum_{\nu=\pm 1}\left[ \left(u_++\nu u_-\right)\ln\left(u_++\nu u_-\right)+
\left(v_++\nu v_-\right)\ln\left(v_++\nu v_-\right)\right.\right.\nonumber\\
&&\left.\left.+\left(r_++\nu r_-\right)\ln\left(r_++\nu r_-\right)\right]\right\},
\end{eqnarray}
while the excess part has the same expression (\ref{exceso}). 
Minimizing the total free energy density $\Phi^*_{\rm id}+\Phi^*_{\rm exc}$ with respect to 
$u_{\pm}$, $v_{\pm}$ and $r_{\pm}$, we obtain 
\begin{eqnarray}
&&\ln(u_+^2-u_-^2)+2y\left[1+y(s_+^2-s_-^2)\right]\kappa_2+2y s_+(\kappa_2+1)+2\Lambda=0, \label{d}\\
&&\ln(v_+^2-v_-^2)+2y\left[1+y(s_+^2-s_-^2)\right]\kappa_1+2y s_+(\kappa_1+1)+2\Lambda=0,\label{e}\\
&&\ln(r_+^2-r_-^2)+2y\left[1+y(s_+^2-s_-^2)\right]\kappa_1\kappa_2+2y s_+(\kappa_1+\kappa_2)+2\Lambda=0,\label{f}\\
&&\ln\left(\frac{u_++u_-}{u_+-u_-}\right)-2y s_-(\kappa_2-1)=0, \label{a}\\
&&\ln\left(\frac{v_++v_-}{v_+-v_-}\right)-2y s_-(\kappa_1-1)=0, \label{b}\\
&&\ln\left(\frac{r_++r_-}{r_+-r_-}\right)-2y s_-(\kappa_1-\kappa_2)=0,\label{c} 
\end{eqnarray}
where $\Lambda$ is a Lagrange multiplier which guarantees the constraint $\displaystyle{2\left(u_++v_++r_+\right)=1}$. 
Considering the case of vanishingly small biaxial ordering, i.e. $u_-\sim 0$, $v_-\sim 0$ and $r_-\sim 0$ 
(which is correct near and above the bifurcation point), we can expand 
Eqns. (\ref{a}), (\ref{b}) and (\ref{c}) up to first order in these variables to obtain in matrix form
$A\cdot {\bm h}={\bf 0}$, where we have defined the vector ${\bm h}^T=(u_-,v_-,r_-)$ and a matrix $A$ with 
the form
\begin{eqnarray}
A=\begin{pmatrix}
1-y u_+(\kappa_2-1)^2 & -y u_+(\kappa_1-1)(\kappa_2-1) & -y u_+(\kappa_1-\kappa_2)(\kappa_2-1)\\
-y v_+(\kappa_1-1)(\kappa_2-1) & 1-y v_+(\kappa_1-1)^2 & -y v_+(\kappa_1-\kappa_2)(\kappa_1-1) \\
-y r_+(\kappa_1-\kappa_2)(\kappa_2-1) & -y r_+(\kappa_1-\kappa_2)(\kappa_1-1) & 1-y r_+(\kappa_1-\kappa_2)^2,
\end{pmatrix}
\end{eqnarray}
This matrix is to be evaluated at $u_+=\gamma_{zx}$, $v_+=\gamma_{xz}$ and $r_+=\gamma_{xy}$ (the values for 
uniaxial symmetry). A nontrivial solution 
of $A\cdot {\bm h}={\bf 0}$ is obtained when $\text{det}\left(A\right)=0$, which is equivalent to the condition
\begin{eqnarray}
y^{-1}&=&u_+(\kappa_2-1)^2+v_+(\kappa_1-1)^2+r_+(\kappa_1-\kappa_2)^2\nonumber\\&=&
\frac{(\kappa_1-\kappa_2)^2}{2}-(\kappa_1-1)(\kappa_1+1-2\kappa_2)u_+-(\kappa_2-1)(\kappa_2+1-2\kappa_1)v_+.
\end{eqnarray}
The values of $u_+$, $v_+$ and $r_+$ at the bifurcation are those obtained from (\ref{d}), (\ref{e}) 
and (\ref{f}) taking 
$u_-=v_-=r_-=0$. Note that, as they are not independent variables, we can solve only the equations for $u_+$ and 
$v_+$, and substitute $r_+=1/2-u_+-v_+$ in all the parameters depending on $r_+$. The result is:
\begin{eqnarray}
&&f_1(u_+,v_+)\equiv u_+-C^{-1}(u_+,v_+) e^{-\xi_1(u_+,v_+)}=0,\nonumber \\
&&f_2(u_+,v_+)\equiv v_+-C^{-1}(u_+,v_+) e^{-\xi_2(u_+,v_+)}=0,\nonumber \\
&&C(u_+,v_+)=2\left[e^{-\xi_1(u_+,v_+)}+e^{-\xi_2(u_+,v_+)}+e^{-\xi_3(u_+,v_+)}\right],
\label{set}
\end{eqnarray}
where we have defined 
\begin{eqnarray}
&&\xi_1(u_+,v_+)=y\left[\kappa_2(1+y s_+^2)+(\kappa_2+1)s_+\right], \\
&&\xi_2(u_+,v_+)=y\left[\kappa_1(1+y s_+^2)+(\kappa_1+1)s_+\right], \\
&&\xi_3(u_+,v_+)=y\left[\kappa_1\kappa_2(1+y s_+^2)+(\kappa_1+\kappa_2)s_+\right],
\end{eqnarray}
and it is convenient to rewrite $s_+$, considering that $u_++v_++r_+=1/2$, as 
\begin{eqnarray}
s_+=\frac{\kappa_1+\kappa_2}{2}-(\kappa_1-1)u_+-(\kappa_2-1)v_+.
\end{eqnarray}
Once the values of $u_+$ and $v_+$ are found by solving the set (\ref{set}), the packing fraction 
at which the bifurcation occurs can be calculated from 
\begin{eqnarray}
\eta=\frac{y \Psi_2}{1+y\Psi_2},\quad \Psi_2=\kappa_1\kappa_2-2\kappa_2(\kappa_1-1)u_+-2\kappa_1(\kappa_2-1)v_+.
\label{packing}
\end{eqnarray}
The set of end-points separating the regions in the $\kappa_1-\kappa_2$ plane where the system (\ref{set})
has a different number of solutions can be calculated by equating the Jacobian to zero:
\begin{eqnarray}
J(u_+,v_+)\equiv \left|
\begin{matrix}
\displaystyle{\frac{\partial f_1}{\partial u_+}} & \displaystyle{\frac{\partial f_1}{\partial v_+}}\\\\
\displaystyle{\frac{\partial f_2}{\partial u_+}} & \displaystyle{\frac{\partial f_2}{\partial v_+}}
\end{matrix}
\right|.
\end{eqnarray} 
Using Eqs. (\ref{set}) we obtain:  
\begin{eqnarray}
J(u_+,v_+)&=&1+3u_+v_+(1-2u_+-2v_+)y^3(\kappa_1-1)^2(\kappa_2-1)^2(\kappa_2-\kappa_1)^2\nonumber\\
&+&u_+(1-2u_+)\frac{\partial\xi_{13}}{\partial u_+}+v_+(1-2v_+)\frac{\partial \xi_{23}}{\partial v_+}
-2u_+v_+\left(\frac{\partial\xi_{23}}{\partial u_+}+\frac{\partial\xi_{13}}{\partial v_+}\right),
\label{end}
\end{eqnarray}
where the explicit expressions for the functions $\partial \xi_{i3}/\partial (u_+,v_+)$ ($\xi_{i3}=\xi_i-\xi_3$ and $i=1,2$) are:
\begin{eqnarray}
&&\frac{\partial\xi_{13}}{\partial u_+}=-y(\kappa_1-1)^2\left[y\left(\kappa_2+(1+2y\kappa_2s_+)s_+\right)
(\kappa_1+1-2\kappa_2)-1-2y\kappa_2s_+\right],\nonumber\\
&&\frac{\partial\xi_{13}}{\partial v_+}=-y(\kappa_1-1)(\kappa_2-1)\left[y\left(\kappa_2+(1+2y\kappa_2s_+)s_+\right)
(\kappa_2+1-2\kappa_1)-1-2y\kappa_2s_+\right],\nonumber\\
&&\frac{\partial\xi_{23}}{\partial u_+}=-y(\kappa_1-1)(\kappa_2-1)\left[y\left(\kappa_1+(1+2y\kappa_1s_+)s_+\right)
(\kappa_1+1-2\kappa_2)-1-2y\kappa_1s_+\right],\nonumber\\
&&\frac{\partial\xi_{23}}{\partial v_+}=-y(\kappa_2-1)^2\left[y\left(\kappa_1+(1+2y\kappa_1s_+)s_+\right)
(\kappa_2+1-2\kappa_1)-1-2\kappa_1s_+\right].\nonumber\\
\end{eqnarray}
To compute the values of $u_+$, $v_+$ and $\kappa_1$ (we fix the value of $\kappa_2$) 
for the location of the critical end-point of the N$_u$-N$_b$ transition, we need to solve
Eqs. (\ref{set}) and also the equation: 
\begin{eqnarray}
J(u_+,v_+)=0.   
\label{cep}
\end{eqnarray}

\section{Spinodal instability to nonuniform phases}
\label{apen3}

The spinodal instability of a uniform phase with respect to density modulations of a given symmetry can be 
calculated by searching the singularities of the structure factor matrix, whose elements can be calculated as
\begin{eqnarray}
T_{\mu\nu,\tau\iota}({\bm q},\rho)=\delta_{\mu\nu,\tau\iota}-\rho \sqrt{\gamma_{\mu\nu}\gamma_{\tau\iota}}
\hat{c}_{\mu\nu,\tau\iota}({\bm q},\rho),
\label{elements}
\end{eqnarray}
with $\hat{c}_{\mu\nu,\tau\iota}({\bm q},\rho)$ 
the Fourier transforms of the direct correlation functions, calculated from the second 
functional derivatives of ${\cal F}_{\rm exc}[\{\rho_{\mu\nu}\}]$ with respect to density profiles. The latter can be computed as
\begin{eqnarray}
-\hat{c}_{\mu\nu,\tau\iota}({\bm q},\rho)=\sum_{\alpha,\beta}\frac{\partial^2\Phi_{\rm exc}}{\partial n_{\alpha}\partial n_{\beta}}
\hat{\omega}^{(\alpha)}_{\mu\nu}({\bm q})\hat{\omega}^{(\beta)}_{\tau\iota}({\bm q}),
\end{eqnarray}
where the Fourier transforms of the weighting functions are
\begin{eqnarray}
&&\hat{\omega}^{(0)}_{\mu\nu}({\bm q})=\hat{w}^{(0)}_{\mu\nu}({\bm q})=\prod_{\tau=x,y} \chi_0 (q_{\tau}^*\kappa^{\tau}_{\mu\nu}/2),\\
&&\hat{\omega}^{(2)}_{\mu\nu}({\bm q})=\sigma_3^2\hat{w}^{(2)}_{\mu\nu}({\bm q})=
\sigma_3^2\prod_{\tau=x,y}\kappa_{\mu\nu}^{\tau}\chi_1 (q_{\tau}^*\kappa^{\tau}_{\mu\nu}/2),\\
&&\hat{\omega}^{(1x)}_{\mu\nu}({\bm q})=\sigma_3\hat{w}^{(1x)}_{\mu\nu}({\bm q})=
\sigma_3\kappa_{\mu\nu}^x\chi_1 (q_x^*\kappa^{x}_{\mu\nu}/2)\chi_0(q_y^*\kappa^y_{\mu\nu}/2),\\
&&\hat{\omega}^{(1y)}_{\mu\nu}({\bm q})=\sigma_3\hat{w}^{(1y)}_{\mu\nu}({\bm q})
=\sigma_3\kappa_{\mu\nu}^y\chi_0(q_x^*\kappa^x_{\mu\nu}/2)\chi_1(q_y^*\kappa^{y}_{\mu\nu}/2),
\end{eqnarray}
(with $\chi_0(x)=\cos x$, $\chi_1(x)=\sin(x)/x$ for $x\neq 0$ while $\chi_1(0)=1$, and $q_{\tau}^*=q_{\tau}\sigma_3$).

The elements (\ref{elements}) can be written in the following explicit form:
\begin{eqnarray}
&&T_{\mu\nu,\tau\iota}=\delta_{\mu\nu,\tau\iota}+y\sqrt{\gamma_{\mu\nu}\gamma_{\tau\iota}}\left\{
\langle\hat{w}^{(0)}_{\mu\nu}({\bm q}^*)\hat{w}^{(2)}_{\tau\iota}({\bm q}^*)\rangle
+\langle\hat{w}^{(1x)}_{\mu\nu}({\bm q}^*)\hat{w}^{(1y)}_{\tau\iota}({\bm q}^*)\rangle\right.\\
&&\left.+y\left[\Psi_{1y}\langle \hat{w}^{(1x)}_{\mu\nu}({\bm q}^*)\hat{w}^{(2)}_{\tau\iota}({\bm q}^*)\rangle
+\Psi_{1x}\langle \hat{w}^{(1y)}_{\mu\nu}({\bm q}^*)\hat{w}^{(2)}_{\tau\iota}({\bm q}^*)\rangle
+\left(1+2y\Psi_{1x}\Psi_{1y}\right)\hat{w}^{(2)}_{\mu\nu}({\bm q}^*)\hat{w}^{(2)}_{\tau\iota}({\bm q}^*)\right]\right\},
\nonumber\\
\label{explicit}
\end{eqnarray}
where we have defined 
\begin{eqnarray}
\langle \hat{w}^{(\alpha)}_{\mu\nu}({\bm q}^*)\hat{w}^{(\beta)}_{\tau\iota}({\bm q}^*)\rangle
= \hat{w}^{(\alpha)}_{\mu\nu}({\bm q}^*)\hat{w}^{(\beta)}_{\tau\iota}({\bm q}^*)+
\hat{w}^{(\beta)}_{\mu\nu}({\bm q}^*)\hat{w}^{(\alpha)}_{\tau\iota}({\bm q}^*).
\end{eqnarray}
Therefore the spinodal instability of a uniform phase with respect to density modulations can be found from 
\begin{eqnarray}
|T({\bm q}^*,\rho^*)|=0,\label{determina}
\label{non_uniform}
\end{eqnarray}
where $|T({\bm q}^*,\rho^*)|$ denotes the determinant of the $6\times 6$ symmetric matrix with elements given
by (\ref{explicit}). In this way 
we find the values $\rho^*_b$ and ${\bm q}^*_b$ at the bifurcation for which the absolute minimum of 
$|T({\bm q}^*,\rho^*)|$ as a function of ${\bm q}^*$ is equal to zero for the first time.
In practice we select ${\bm q}^*=(q_x^*,0)$ or ${\bm q}^*=(0,q_y^*)$ with $q_{x,y}^*=2\pi\sigma_3/d_{x,y}$ where $d_{x}$ and $d_{y}$ 
are the periods of nonuniform phases along $x$ and $y$ respectively. 
The values $\{\gamma_{\mu\nu}\}$ at each step of the 
numerical procedure used to solve Eqns. (\ref{determina}) are found from the solution of Eqn. (\ref{itera}). Bifurcated phases can have 
different symmetries: smectic (S) or columnar (C), where density modulations exist along only one 
spatial direction which could coincide 
(S) or not (C) with the alignment directions of the particle projections. Also a crystalline phase (K) with 
full 2D positional ordering could exist with (orientational ordered K) or without (plastic K) orientational 
ordering.

\end{document}